\documentclass[aps,prb,reprint,superscriptaddress]{revtex4-2}
\usepackage{graphicx}
\usepackage{latexsym}
\usepackage{amssymb}
\usepackage{amsmath}
\usepackage{amsfonts}
\usepackage{lipsum} 
\usepackage[vcentermath]{youngtab}
\usepackage{upgreek}
\usepackage{bm,dsfont}
\usepackage{multirow}
\usepackage{enumitem}
\usepackage[usenames, dvipsnames]{color}
\usepackage[svgnames]{xcolor}
\usepackage{hyperref}
\hypersetup{
colorlinks = true,
linkcolor = [rgb]{0.70,0.13,0.13},
citecolor = [rgb]{0.13,0.55,0.13},
urlcolor = [rgb]{0.25, 0.41, 0.88}}
\usepackage{makecell}
\usepackage{ulem}
\newcommand{\bra}[1]{{\langle #1|}}
\newcommand{\ket}[1]{{|#1 \rangle}}

\newcommand{\dd}{\mathrm{d}}
\newcommand{\ii}{\mathrm{i}}

\renewcommand{\O}{\mathrm{O}}

\newcommand{\sgn}{\mathop{\operatorname{sgn}}}

\newcommand{\vect}[1]{{\bm{#1}}}

\newcommand{\eqnref}[1]{Eq.\,\eqref{#1}}
\newcommand{\figref}[1]{Fig.\,\ref{#1}}
\newcommand{\tabref}[1]{Tab.\,\ref{#1}}

\begin{document}

\title{Symmetric Mass Generation of K\"ahler-Dirac Fermions from the Perspective of Symmetry-Protected Topological Phases 
}

\author{Yuxuan Guo}
\affiliation{School of Physics, Peking University, Beijing 100871, China}
\author{Yi-Zhuang You}
\affiliation{Department of Physics, University of California San Diego, La Jolla, CA 92093, USA}
\begin{abstract}
The K\"ahler-Dirac fermion, recognized as an elegant geometric approach, offers an alternative to traditional representations of relativistic fermions. Recent studies have demonstrated that symmetric mass generation (SMG) can precisely occur with two copies of K\"ahler-Dirac fermions across any spacetime dimensions. This conclusion stems from the study of anomaly cancellation within the fermion system.
Our research provides an alternative understanding of this phenomenon from a condensed matter perspective, by associating the interacting K\"ahler-Dirac fermion with the boundary of bosonic symmetry-protected topological (SPT) phases. We show that the low-energy bosonic fluctuations in a single copy of the K\"ahler-Dirac fermion can be mapped to the boundary modes of a $\mathbb{Z}_2$-classified bosonic SPT state, protected by an inversion symmetry universally across all dimensions.
This implies that two copies of K\"ahler-Dirac fermions can always undergo SMG through interactions mediated by these bosonic modes. This picture aids in systematically designing SMG interactions for K\"ahler-Dirac fermions in any dimension. We present the exact lattice Hamiltonian of these interactions and validate their efficacy in driving SMG.

\end{abstract}
\maketitle
\date{\today}

\section{Introduction}
Symmetric mass generation (SMG) \cite{Fidkowski:2010bf,Fidkowski:2011dd,Wang2013Non-Perturbative,Slagle:2015lo,Ayyar2015Massive,Catterall:2016sw,Tong2021Comments,Wang2204.14271} is a novel mechanism for gapless fermions to acquire a many-body excitation gap without spontaneous symmetry breaking. It occurs in a fermion system that is (i) free of any quantum anomaly \cite{Ryu2012Electromagnetic,Kapustin2014Anomalies,Tanizaki2018Anomaly,Tachikawa1805.02772,Yamaguchi2019t-Hooft}, ensuring no obstruction towards trivial gapping; yet (ii) the symmetry is restrictive enough to rule out any fermion bilinear mass, thus gapping can only be achieved by multi-fermion condensation.

The SMG mechanism has attracted much research interest in both high-energy and condensed matter physics \cite{Ayyar:2016fi,Catterall:2016nh,Ayyar:2016tg,Witten:2016yb,Ayyar:2016ph,He:2016qy,Ayyar2017Generating,You2018Symmetric,You2018From,Schaich2018Phases,Catterall2018Topology,Butt2018Four,Butt2018SO4-invariant,Catterall2020Exotic,Xu2021Greens,Catterall2021Chiral,Butt2021Symmetric,Lu2210.16304}. In condensed matter physics, the study of SMG was initially driven by the classification problem of interacting fermionic symmetry-protected topological (SPT) phases \cite{Gu2012Symmetry-protected,Cheng2015Classification,Morimoto1505.06341,Kapustin2015Fermionic,Freed2016Reflection,Gaiotto2016Spin,WenZoo1610.03911,Wang1703.10937,Kapustin2017Fermionic,Wang2018Tunneling,Wang2018Construction,Gaiotto2019Symmetry,Tong1906.07199,Lan2019Fermion,Guo2020Fermionic,Ouyang2020Computing,Aasen2109.10911,Barkeshli2109.11039,Hasenfratz2204.04801,Manjunath2210.02452,Zhang2211.09127,Zhang2204.05320}. Some fermionic SPT classifications reduce from $\mathbb{Z}$ to $\mathbb{Z}_N$ in the presence of interactions \cite{Fidkowski:2010bf,Fidkowski:2011dd,Ryu:2012ph,Qi:2013qe,Yao:2013yg,Wang2014Interacting,Gu:2014tw,Metlitski2014Interaction,You:2014ho,Yoshida:2015aj,Gu:2015cy,Song1609.07469,Queiroz:2016se}. Consequently, the gapless fermion edge states of $N$ copies of interacting fermionic SPT states can be trivially gapped by the SMG mechanism without condensing any symmetry-breaking fermionic bilinear mass. In high-energy physics, the SMG mechanism was originally proposed as a solution to regularize chiral fermions on the lattice \cite{Wang2013Non-Perturbative,Wen:2013kr,You:2014ow,You:2015lj,BenTov:2015lh,DeMarco2017A-Novel,Wang2018A-Non-Perturbative,Wang2019Solution,Kikukawa2019Why-is-the-mission,Razamat2021Gapped,Butt2021Anomalies,Zeng2022Symmetric}, circumventing the Nielsen-Ninomiya theorem \cite{Nielsen1981Absence,Nielsen:1981we,Nielsen:1981bc} and solving the fermion doubling problem by introducing non-perturbative interaction effects.

The SMG can occur in Majorana fermions in any spacetime dimension \cite{You:2014ho}. One intriguing feature is that it only occurs when the fermion flavor number is a multiple of a certain "magic number" $N$. This number $N$ is determined by the anomaly cancellation condition (or the gapping condition) and varies with the spacetime dimension in a rather complicated pattern as enumerated in \tabref{tab: magic number} for Majorana/Weyl fermions. However, it was recently discovered \cite{Butt2021Anomalies,Catterall2201.00750,Catterall2209.03828} that the SMG of K\"ahler-Dirac fermions \cite{Benn1983fermions,SimonPRD} enjoys a strikingly simple flavor number requirement: every two copies of K\"ahler-Dirac fermions can always be gapped by SMG regardless of the spacetime dimension. The K\"ahler-Dirac fermion provides an alternative representation of gapless relativistic fermions, utilizing antisymmetric tensor fields as opposed to traditional spinor fields. This approach was initially conceived as an extension of the Kogut-Susskind staggered fermion \cite{susskind1976}. It possesses an elegant geometric construction on the lattice and can be systematically defined in the flat spacetime of any dimension.

\begin{table}[htp]
\caption{The required fermion flavor number $N$ in each spacetime dimension $D$ for SMG to occur.}
\begin{center}
\begin{tabular}{c|cccccccc}
$D=d+1$ & 1 & 2 & 3 & 4 & 5 & 6 & 7 & 8\\
\hline
Majorana& 8 & 8 & 16 & 16 & 32 & 32 & 64 & 64 \\
K\"ahler-Dirac & 2 & 2 & 2 & 2 & 2 & 2 & 2 & 2
\end{tabular}
\end{center}
\label{tab: magic number}
\end{table}

The goal of this work is to explore the deeper reason behind the universal flavor number requirement for the SMG of K\"ahler-Dirac fermions. Although this flavor number requirement can be (and has been) obtained through anomaly cancellation analysis \cite{Butt2021Anomalies,Catterall2201.00750,Catterall2209.03828}, here we would like to provide a more condensed-matter-oriented explanation. We first establish the connection that a single copy of interacting K\"ahler-Dirac fermion in any spacetime dimension $D$ can be mapped to an $\O(D+1)$ sigma model (NL$\sigma$M) with a Wess-Zumino-Witten (WZW) term, describing the boundary of a bosonic SPT phase \cite{Bi:2015qv,You:2014ho} in the $D$-dimensional bulk. Then, we show that this bulk bosonic SPT phase is universally $\mathbb{Z}_2$-classified under the spatial inversion symmetry $\mathbb{Z}_2^P$, regardless of the spacetime dimension $D$. Therefore, every two copies of K\"ahler-Dirac fermion correspond to the boundary of a trivial SPT phase, which has no obstruction towards trivial gapping, thus allowing SMG to occur.

Our method offers a unified theoretical framework for SMG of Majorana and Weyl fermions across all spatial dimensions, apart from the calculation of the anomaly-free condition. Furthermore, we provide an extensive exploration of the second quantization of the K\"ahler-Dirac equation, illustrating how fermions hop on the lattice and giving a very simple geometric interpretation to it.

\section{Lattice Model}

\subsection{Spacetime Lattice and Fermionic Fields}

\begin{figure}[htbp]
\begin{center}
\includegraphics[scale=0.65]{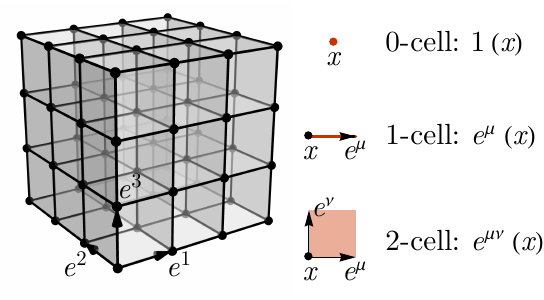}
\caption{A hypercubic lattice in $D=3$ dimensional flat spacetime, with a few examples of $p$-simplices. In fact, the cells aren't triangles; the simplex refers to a singular simplex in mathematics.}
\label{fig: lattice}
\end{center}
\end{figure}

As illustrated in \figref{fig: lattice}, the K\"ahler-Dirac fermion in flat spacetime can be defined on a hypercubic lattice equipped with the metric $\eta^{\mu\nu}$, where $\eta^{\mu\nu}$ could correspond to either the Euclidean or Minkowski metric. A $D$-dimensional hypercubic lattice is defined by a set of orthonormal lattice vectors $e^{\mu}$ ($\mu=0,1,2,\ldots,d=D-1$) with an inner product structure $e^{\mu}\cdot e^{\nu}=-\eta^{\mu \nu}$, such that each lattice point can be represented as a linear combination of these lattice vectors with integer coefficients, given by $x=\sum_{\mu=1}^{D} m_\mu e^\mu$ with $m_\mu\in\mathbb{Z}$. Unlike conventional lattice fermion models, where fermion modes are defined on lattice points (0-chains) exclusively, in the lattice model of K\"ahler-Dirac fermions, each cell on the lattice is associated with a fermion mode. This approach avoids the doubling problem, ensuring that in the continuum limit of the lattice model, wherein fermionic fields are defined on forms, the same number of copies of Dirac fermions is retained.

On the hypercubic lattice, the co-chain is isomorphic to chain, and we denote a basis of co-chain as $e^{\bm{\mu}}(x)$ as the dual of $e_{\bm{\mu}}(x)$, which satisfies $e^{x,\bm{\mu}}(e_{x',\bm{\mu}'})=\delta_{x,x'}\delta_{\bm{\mu},\bm{\mu}'}$

The cell centers form a refined lattice with each unit cell containing $2^D$ points. The fermion field $\psi_\vect{\mu}(x)$ transforms as a rank-$p$ anti-symmetric tensor under point group rotations of the lattice. The field $\psi_\vect{\mu}(x)$ can pair up with the $p$-cell $e^\vect{\mu}(x)$ to define the following fermionic field on the lattice:
\begin{equation}\label{eq: Psi}
\Psi(x) = \sum_{\vect{\mu}}\psi_\vect{\mu}(x)e^\vect{\mu}(x),
\end{equation}
which collects all fermion fields on the $2^D$ cells within the unit cell labeled by $x$. As a result, $\Psi(x)$ has $2^D$ complex components correspondingly.

Similar to differential forms, we can define the fundamental operations of the exterior product $\wedge$ and the interior product $\iota$ for cells. The exterior product $\wedge$ is defined as
\begin{gather}
e^{\bm{\mu}}(x) \wedge e^{\bm{\mu}'}(x') = {\sgn(\bm{\mu}\cup\bm{\mu}')} \delta(x-x') e^{\bm{\mu} \cup\bm{\mu}'}(x),
\end{gather}
where $\bm{\mu}\cup\bm{\mu}'$ denotes the union of the two index sets $\bm{\mu}$ and $\bm{\mu}'$, sorted in ascending order. $\sgn(\bm{\mu}\cup\bm{\mu}')$ represents the sign of the permutation required to sort the sequence $\mu_1\mu_2\cdots\mu_p\mu'_1\cdots\mu'_p$ into ascending order. If $\vect{\mu}$ and $\vect{\mu}'$ share at least one common index, $\sgn(\bm{\mu}\cup\bm{\mu}')=0$ is assigned.

The interior product $\iota$ is defined similarly
\begin{gather}
    e^{\bm{\mu}}(x)\iota e^{\bm{\mu}'}(x')=\text{sgn}(\bm{\mu}/\bm{\mu}')\delta(x-x') e^{\overline{\bm{\mu}}}(x),
\end{gather}

We assume $\bm{\mu}\subset\bm{\mu}'$ and $\text{sgn}(\bm{\mu}/\bm{\mu}')=(-1)^p\sgn({\bm{\mu}\overline{\bm{\mu}}})\sgn({\bm{\mu}'})$, where $\overline{\bm{\mu}}$ only contains elements doesn't belong to $\bm{\mu}$ but belongs to $\bm{\mu}'$. Indeed, from a mathematical perspective, a lattice constitutes a particular type of chain, which makes the rigorous definition of interior product $\iota$ and exterior product $\wedge$ on a general lattice challenging and the true definition is cap product $\cap$ and cup  product $\cup $ (details could be found in appendix.\ref{appendix:pro}). However, for hypercubic lattices, the computational rules remain consistent, enabling the straightforward use of exterior and interior product notations. This specificity streamlines the mathematical manipulations and contributes to the practical convenience of our model. 

\subsection{Lagrangian Formulation}
\label{sec2B}
In this section, we discuss the operations of K\"ahler-Dirac fermions on a hypercubic lattice, as presented in the related literature \cite{r9}.

The K\"ahler-Dirac operator, as a type of Dirac operator, acts on differential forms on the spacetime manifold as a "square root" of the Laplace operator $\Delta = \partial^\mu\partial_\mu$. In continuous spacetime, one can define the exterior derivative $\dd = \partial_\mu \dd x^\mu\wedge$ and its adjoint $\delta \equiv \dd^\dagger = \partial_\mu \dd x^\mu \iota$. They satisfy $\dd^2 = \delta^2 = 0$ and are related by the Hodge dual as $\delta = (-1)^p \star^{-1} \dd \star$ when acting on $p$-forms. The K\"ahler-Dirac operator $K$ is simply defined as
\begin{equation}\label{eq: def K diff}
K = \dd - \delta = \partial_\mu \dd x^\mu\wedge - \partial_\mu\dd x^\mu \iota,
\end{equation}
such that $K^2 = (d - \delta)^2 = \star\dd\star\dd + \dd\star\dd\star = 2\Delta$, where $\Delta$ is the Laplace operator. This is consistent with the property that the square of the Dirac operator produces the Laplace operator.

To define the K\"ahler-Dirac operator on the hypercubic lattice, two lattice difference operators should be introduced: the forward difference $\partial_\mu^+ = T(e^{\mu}) - 1$ and the backward difference $\partial_\mu^- = -(\partial_\mu^+)^\dagger = 1 - T(-e^{\mu})$, where $T(e^\mu)$ denotes the translation operator that translates the field it acts on by the lattice vector $e^\mu$. More explicitly,
\begin{align}
\partial_\mu^+\psi_{\vect{\nu}}(x) &= \psi_{\vect{\nu}}(x + e^\mu) - \psi_{\vect{\nu}}(x),\\
\partial_\mu^-\psi_{\vect{\nu}}(x) &= \psi_{\vect{\nu}}(x) - \psi_{\vect{\nu}}(x - e^\mu).
\end{align}
With this notation, the lattice K\"ahler-Dirac operator $K$ can be expressed as
\begin{equation}
K = \partial_{\mu}^- e^{\mu}\wedge - \partial_{\mu}^+ e^{\mu}\iota.
\end{equation}
The lattice K\"ahler-Dirac operator $K$ approaches its continuum limit in \eqnref{eq: def K diff} when $e^{\mu}\rightarrow dx^{\mu}$ and $\partial_{\mu}^{\pm}\rightarrow \partial_{\mu}$.

The lattice model of K\"ahler-Dirac fermions is then described by the following action
\begin{equation}
S = \sum_{x}\overline{\Psi} (\partial_{\mu}^-e^{\mu}\wedge - \partial_{\mu}^+e^{\mu}\iota - m)\Psi,
\end{equation}
where $\Psi = \Psi_{\vect{\mu}}e^{\vect{\mu}}$ is the fermion field defined in \eqnref{eq: Psi}. $\overline{\Psi} = \Psi^{\dagger}e^0(\wedge - \iota)$ symbolizes the conjugation of $\Psi$, and $\overline{\Psi}$ is an independent field in the action. Taking the saddle point equation $\delta S/\delta \overline{\Psi} = 0$ yields the K\"ahler-Dirac equation for fermions on a lattice:
\begin{equation}
(\partial_{\mu}^-e^{\mu}\wedge - \partial_{\mu}^+e^{\mu}\iota - m)\Psi(x) = 0,
\end{equation}
where $m$ parameterizes the fermion mass. This describes $2^{\lceil D/2 \rceil}$ copies of Dirac fermions in flat spacetime, where $\lceil x\rceil$ represents the ceiling function. For instance, when $d+1=4$, and the dimension of the Hilbert space, $dim(\mathcal{H})=2^{4}=16$, can be decomposed into invariant subspaces $\bigoplus_{a=1}^4\mathcal{V}^{a}$. The K\"ahler-Dirac equation implies $(i\partial_{\mu}\gamma^{\mu} - m)\Phi^a = 0, \Phi^a\in\mathcal{V}^a$, which suggests that K\"ahler-Dirac fermions correspond to four copies of Dirac fermions in $(3+1)d$ \cite{r12}.

The analogy of exterior derivative $\dd$ and its conjugation on lattce are $
\dd=\partial^-_\mu e^{\mu}\wedge\nonumber$ and $\delta=\partial^+_\mu e^{\mu}\iota$. In the context of the continuum limit, $d$ and $\delta$ correspond to nothing other than the differential operator $d$ and co-differential $\delta=(-1)^p\star^{-1}d\star$ acting on $p$-forms. By employing these operators, the action of Kähler-Dirac fermions can be formulated in a more geometric manner:

\begin{gather}
S=\int d^Dx\overline{\Psi}(d-\delta-m)\Psi
\end{gather}

\subsection{Halmitonian Picture}
We now provide a Hamiltonian formulation for Kähler-Dirac fermions following the process of second quantization.

At the single-particle level, we can formulate a Dirac-like equation similar to the Kähler-Dirac equation up to basis transformations. To obtain a Hamiltonian picture, we need to take a continuum limit in the time direction and keep the space components discrete in Minkowski spacetime. The equation of motion can be expressed as follows:
\begin{equation}
\ii\partial_0\Psi(x) = \ii(\partial_{i}^-e^{i}\wedge - \partial_{i}^+e^{i}\iota + m e^0(\wedge - \iota))\Psi(x),
\end{equation}
where $i \neq 0$.

An equivalent form of the equation of motion is
\begin{equation}
\ii\partial_0\Psi(x) = (\ii(\dd' - \delta') + \ii m \dd x^0(\wedge - \iota))\Psi(x).
\end{equation}
In these equations, $d'$ and $\delta'$ represent the restrictions of $d$ and $\delta$ on $d$-dimensional space, respectively. Here, $\dd' e^{x;\bm{\mu}} = \sum_{i=1}^{d}\text{sgn}({i\cup \bm{\mu}})(e^{x-e_i;i\cup \bm{\mu}} - e^{x;i\cup \bm{\mu}})$ and $\delta' e^{x;\bm{\mu}} = \sum_{i=1}^{d}\text{sgn}(i/\bm{\mu})(e^{x;i/\bm{\mu}} - e^{x+e_{i};i/\bm{\mu}})$. The strength of this formulation lies in its capacity to maintain simplicity in the geometric interpretation of space components. Since we are interested in the case when the physical mass $m=0$ where SMG can happen, the Hamiltonian is further simplified as $H = i(d' - \delta')$. The second quantization of this Hamiltonian gives a tight-binding-like model:
\begin{equation}
H = \sum_{x,\bm{\mu}}\left(\sum_{i}\ii\sgn(i/\bm{\mu})(\psi^\dagger_{x;i/\bm{\mu}} - \psi^\dagger_{x+e^i;i/\bm{\mu}})\psi_{x;\bm{\mu}} + \text{h.c.}\right),
\end{equation}
where the first summation $\sum_{x;\bm{\mu}}$ runs over all simplexes on an equal time slice, and the second summation $\sum_{i}$ runs over the basis of space $e^{i}$. The annihilation operator is denoted as $\psi_{x;e^{\bm{\mu}}}$, which nullifies fermions existing on the cell $e^{\bm{\mu}}(x)$. The annihilation and creation operators satisfy $\{\psi^\dagger_{x;\bm{\mu}} ,\psi_{x';\bm{\mu}'}\}=\delta(x-x') \delta_{\bm{\mu},\bm{\mu}'}$, where $x$ and $x'$ belong to the same equal time slice. After the second quantization, each simplex is attributed with a fermion.

In order to simplify this equation, we define the annihilation operator on a chain rather than on a simplex\footnote{A chain is the sum of several simplexes.}, as $\psi_{\sigma_1+\sigma_2} = \psi_{\sigma_1} + \psi_{\sigma_2}$, where $\sigma_1$ and $\sigma_2$ are distinct simplexes. Employing this configuration, we can express the Hamiltonian in a more geometric fashion:
\begin{equation}
H = \ii\sum_{\sigma}(\psi^{\dagger}_{\sigma}\psi_{\partial' \sigma} - \psi^{\dagger}_{\partial' \sigma}\psi_{\sigma}),
\end{equation}
where the summation runs over all simplexes $\sigma$ on an equal time slice, and $\partial'=\sum_{i\neq0}\text{sgn}(i\cup\bm{\mu})(e_{\vect{\mu}}(x+e^i)-e_{\vect{\mu}}(x))$ is the restriction of the boundary on the equal time slice, which may be a sum of different simplexes. It becomes evident that hopping occurs exclusively between a simplex and the restriction of the boundary of that simplex.

\section{NL$\sigma$M from K\"ahler-Dirac Fermions}

In Section \ref{sec2B}, we discuss the inclusion of Kähler-Dirac fermions, which comprise $2^{\lceil D/2 \rceil}$ distinct copies of Dirac fermions in the context of flat spacetime. This inherent redundancy enables the fermionic model to be effectively mapped onto a Nonlinear $\sigma$ Model (NL$\sigma$M) through the coupling with an $O(d+2)$ bosonic field \cite{You:2014ho}. In this section, we demonstrate the procedure for coupling the Kähler-Dirac fermions with the bosonic field, leading to the emergence of an NL$\sigma$M.

The bosonic field plays a crucial role in facilitating interactions among distinct flavors of Kähler-Dirac fermions. Prior studies have established that the flavor symmetry of Kähler-Dirac fermions in Euclidean space corresponds to the conformal symmetry group $SO(1,D+1)$ \cite{R0}, which is large enough to support the NL$\sigma$M.

To determine the additional $(d+2)$ gamma matrices, we will briefly review the origins of the known gamma matrices. These matrices are derived from the square root of the Laplacian:
\begin{equation}
\sqrt{\partial_{\mu}\partial^{\mu}} = \gamma^\mu\partial_{\mu}.
\end{equation}
Next, we consider the square root of $-\partial_{\mu}\partial^{\mu}$, as the square root of an operator is not necessarily linearly dependent on the square root of the negation of the same operator. To be more specific:
\begin{equation}
\sqrt{\Delta} \neq \sqrt{-1}\sqrt{-\Delta}.
\end{equation}
We can verify that $\sqrt{-\Delta} = \dd + \delta$, since $(\dd+\delta)^2 = -(-d\delta-\delta d)$, and we define it as $\Tilde{K} = \dd + \delta$ or the dual Kähler-Dirac operator. Upon choosing a basis, this operator can be decomposed into $i\partial_{\mu}\Tilde{\gamma^{\mu}}$ where $\{\Tilde{\gamma^{\mu}},\Tilde{\gamma^{\nu}}\}=-2\delta^{\mu\nu}$ in Euclidean spacetime. This operator $\tilde{K}$ is linearly independent of $K$ and generates the additional $(d+1)$ gamma matrices. However, we still need to verify whether the new matrices mutually anti-commute with the original gamma matrices generated by $K$. It is straightforward to verify that $\{K,\Tilde{K}\}=\{\dd-\delta,\dd+\delta\}=\{\dd,\delta\}-\{\delta,\dd\}=0$, and since $\partial_{\mu}$ are linearly independent and anti-commute with each other, we can demonstrate that $(\gamma^\mu , \Tilde{\gamma^{\mu}}),\mu=1,2\cdots,d+1$ extend the Clifford algebra to $\mathcal{C}l_{d+1,d+1}$. Similar to the case of $K$ we can represent $\Tilde{K}$ through the exterior product $\wedge$ and the interior product $\iota$:
\begin{equation}
    \Tilde{K} = \partial_{\mu}(\wedge + \iota)dx^{\mu}, \quad \mu=1,2\cdots,d+1.
\end{equation}

However, $\gamma^{\mu}$ and $\Tilde{\gamma^{\mu}}$ do not encompass all matrices that mutually anti-commute. The structure of $\mathcal{C}l_{d+1,d+1}$ enables the definition of one additional chiral operator $\Gamma$ as the highest-grade pseudo-scalar in $\mathcal{C}l_{d+1,d+1}$, which serves as a generator of chiral symmetry and anti-commutes with all gamma matrices. In general, the Clifford algebra, $\Gamma$ is the product of all other independent gamma matrices $\Pi_{\mu}\gamma_{\mu}\Pi_{\mu}\ii\Tilde{\gamma}_{\mu}$, but calculating this product geometrically in our discussion proves to be challenging. Our approach involves identifying a geometric operator that acts on tensors and verifying that it anti-commutes with both $K$ and $\Tilde{K}$. Luckily, the geometric interpretation of $\Gamma$ is quite simple. It turns a $p$-chain into the same $p$-chain up to a sign $(-1)^p$:
\begin{align}
\Gamma : \quad C^p & \longrightarrow C^p \nonumber \\
&e^{\bm{\mu}}(x) \longrightarrow (-1)^{p}e^{\bm{\mu}}(x),
\end{align}
where $\bm{\mu}=\mu_1\mu_2\cdots\mu_p$. In the continuum limit, $\Gamma$ acts on $p$-forms as $\dd x^{\mu_1}\wedge\cdots \dd x^{\mu_p}\rightarrow(-1)^p \dd x^\mu_1\wedge\cdots \dd x^{i_p}$.

$\Gamma$ and $\Tilde{\gamma}^{\mu}$ don't appear in the original K\"ahler-Dirac equation and transfer as a spinor representation of $SO(1,d+1)$.
To couple the Kähler-Dirac field with $n_{\mu}$ and derive the nonlinear sigma model (NL$\sigma$M) on the lattice, we need to replace the translation operator $T(e_\mu)$ in $\Tilde{K}$ with $e^{in_\mu}$ and approximate $e^{in_\mu}-1$ to $in_\mu$. Consequently, we obtain the following equation:
\begin{gather}
    ((\partial^-_{\mu}+i n_{\mu})e^{\mu}\wedge-(\partial^+_{\mu}-i n_{\mu})e^{\mu}\iota+i N\Gamma)\Psi(x)=0
\end{gather}
And the action is:
\begin{gather}
    S=\sum_x \overline{\Psi}(x) ((\partial^-_{\mu}+i n_{\mu})e^{\mu}\wedge-(\partial^+_{\mu}-i n_{\mu})e^{\mu}\iota+i N\Gamma)\Psi(x)
\end{gather}

Also, we could write down the tight binding like model for this fermionic NL$\sigma$M without physical mass term $m$.
\begin{equation}
\begin{split}
\label{eq21}
H=\sum_{x,\bm{\mu}}\Bigg(\Big(\sum_{\mu=0}^d\sum_{i=1}^d\ii(-1)^{i/\bm{\mu}}(\psi^\dagger_{x,i/\bm{\mu}}-\psi^\dagger_{x+e_i,i/\vect{\mu}})\psi_{x,\bm{\mu}}\\+n_\mu (-1)^{\mu/\bm{\mu}} \psi^\dagger_{x,\mu/\bm{\mu}}\psi_{x,\bm{\mu}}+h.c.\Big)+N(-1)^p\psi^\dagger_{x,\bm{\mu}}\psi_{x,\bm{\mu}}\Bigg)
\end{split}
\end{equation}

\section{K\"ahler-Dirac fermions as BSPT}
 $(\Gamma,\Tilde{\gamma_{\mu}})$ transforms as the spinor representation of $O(d+2)$, and the coupling between $(N,n_{\mu})$ and $(\Gamma,\Tilde{\gamma_{\mu}})$ makes the order parameters select a specific direction, and the symmetry group transitions from $O(d+2)$ to $O(d+1)$. This suggests that the effective theory of this model is the quotient space $\frac{O(d+2)}{O(d+1)} \cong S^{d+1}$ NL$\sigma$M. Since $\pi_{d+1}(S^{d+1}) \cong \mathbb{Z}$ is nontrivial, this NL$\sigma$M permits a topological theta term. This analysis can be corroborated by integrating out the fermionic degrees of freedom, and readers can consult reference \cite{r13} for further details regarding the field theory.
\begin{gather}
    \mathcal{S}_{d+1} = \int d^{d+1}x \left( \frac{1}{g^2}\partial_{\mu}n^a\partial^{\mu}n_a + i\frac{\theta}{\Omega_{d+1}}\epsilon_{ab\cdots}n^a\partial_1n^b\partial_2n^c\cdots \right) \\
    \mu=1,2\cdots d+1, \quad a,b,c\cdots=0,1,2\cdots d+1, \quad \theta=2k\pi \nonumber
\end{gather}

Next, we need to determine which symmetry prevents the $\theta$ term from being trivialized. On a lattice, our focus should be directed towards certain discrete subgroups of the orthogonal group. Specifically, the Kähler-Dirac equation possesses an inversion symmetry, denoted as $\mathbb{Z}_{2,P}$, which protects $\mathbb{Z}_2$ SPT phases. The action of the inversion symmetry is represented by the operator $\mathcal{P}=\mathcal{I} e^0(\wedge-\iota)$, where $0$ labels the time direction and $(e^0(\wedge-\iota))^2=-1$, $\mathcal{I}\partial^{\mu}\mathcal{I}^{-1}=-\partial^{\mu}$, thus $\mathcal{P}^2=-1$. The transformation acting on our model can be expressed as follows:
\begin{gather}
\begin{cases}
N \rightarrow -N \\
n_\mu \rightarrow -n_{\mu} \\
\Psi \rightarrow e^0(\wedge-\iota)\Psi
\end{cases}
\end{gather}
In the continuum limit with a certain basis, this symmetry can be represented as $\mathcal{P} = \mathcal{I}dx^0(\wedge-\iota) = \mathcal{I} \gamma^0$. With this understanding in place, we are now prepared to discuss the classification of bosonic SPT phases. It is crucial to note that two copies of Kähler-Dirac fermions can be smoothly connected to a trivial state without breaking any symmetry. This conclusion is substantiated by the existence of an interlayer coupling, given by \cite{Bi:2015qv}:

\begin{gather}
\mathcal{S}_{CP}=-\int d^{D}x AN_1N_2-B n_1^\mu n_{2,\mu}
\end{gather}
where $1,2$ denote different layers. 
Consequently, the effective field theory for the combined field $n$ exhibits a vanishing theta term due to the cancellation of layers.

This coupling of bosonic field can help us derive the SMG interaction of fermions. On the meanfield level, the bosonic field $N\sim\overline{\Psi}\Gamma\Psi$ and $n^{\mu}\sim \overline{\Psi}e^{\mu}(\wedge+\iota)\Psi$. As a result, the interaction of $S_{CP}$ can be ported as the following four fermions interaction:
\begin{equation}
\begin{split}
    S_{CP}=-\int d^{D}x A\overline{\Psi}_1\Gamma\Psi_1\overline{\Psi}_2\Gamma\Psi_2\\-B\overline{\Psi}_1e^{\mu}(\wedge+\iota)\Psi_1\overline{\Psi}_2e^{\mu}(\wedge+\iota)\Psi_2
\end{split}
\end{equation}

In the Hamiltonian picture, we could get the SMG interaction by projecting bosonic field in equation (\ref{eq21}) out.  The coupling can be expressed as:

\begin{gather}
H_{CP}= \int d^dxA\Psi_1^\dagger\Gamma\Psi_1\Psi^\dagger_2\Gamma\Psi_2-B\Psi^\dagger_1e^{\mu}(\wedge+\iota)\Psi_1\Psi^\dagger_2e^{\mu}(\wedge+\iota)\Psi_2\nonumber\\
=\sum_{\bm{\mu},\bm{\mu}',\mu,x}A(-1)^{p+p'}\psi^\dagger_{1,x,\bm{\mu}}\psi_{1,x,\bm{\mu}}\psi^\dagger_{2,x,\bm{\mu}'}\psi_{2,x,\bm{\mu}'}\nonumber\\-B\psi^\dagger_{1,x,\mu/\bm{\mu}}\psi_{1,x,\bm{\mu}}\psi^\dagger_{2,x,\mu/\bm{\mu}'}\psi_{2,x,\bm{\mu}'}
\end{gather}
where $A\sim\langle\Psi_1^\dagger\Gamma\Psi_1\Psi_2^\dagger\Gamma\Psi_2\rangle$ and $B\sim\langle \Psi_1^\dagger e^{\mu}(\wedge+\iota)\Psi_1\Psi_2^\dagger e^{\mu}(\wedge+\iota)\Psi_2\rangle$ come from four fermions condensing and $p,p'$ is the length of $\bm{\mu},\bm{\mu}'$ .
In the strong coupling limit, when $H_{CP}$ donmintes, we can demonstrate that the model possesses a unique and symmetric ground state. We proceed under the assumption that $A\gg B$, treating $B\Psi^\dagger_1e^{\mu}(\wedge+\iota)\Psi_1\Psi^\dagger_2e^{\mu}(\wedge+\iota)\Psi_2$ as a perturbation. It is important to note that $H_{CP}$ does not intermix fields from different $x$ variables, thus it can be understood within the framework of 0+1 dimensional quantum mechanics. The ground state of $H_0=\int d^dxA\Psi^\dagger_1\Gamma\Psi_1\Psi^\dagger_2\Gamma\Psi_2$ is two-fold degenerate, expressed as:
\begin{gather}
\ket{\Psi^+}=\ket{\Psi^1_{\text{even}}}\ket{\Psi^2_{\text{odd}}}\nonumber\\
\ket{\Psi^-}=\ket{\Psi^1_{\text{odd}}}\ket{\Psi^2_{\text{even}}}
\end{gather}
Here, $\ket{\Psi^{1/2}_{even}}=\Pi_{x,\sigma\in\text{even cells}}\psi^\dagger_{\sigma,x}\ket{Vac^{1/2}}$ and $\ket{\Psi^{1/2}_{odd}}=\Pi_{x,\sigma\in\text{ odd cells}}\psi^\dagger_{\sigma,x}\ket{Vac^{1/2}}$, where $\ket{Vac^{1/2}}$ is the vacuum state of sublattice 1 or 2. $\ket{\Psi^{1/2}_{even}}$ and $\ket{\Psi^{1/2}_{odd}}$ label states in which all even or odd simplexes of sublattice 1 or 2 are occupied by fermions.

And since $H_1$ contains hopping from even dimensional simplexes and odd dimensional simplexes. $\bra{\Psi_+}H_1\ket{\Psi_-}\sim B$. The energy degeneracy of $\ket{\Psi_\pm}$ is removed and our model process a unique symmetric ground state:
\begin{gather}
\ket{\Psi_{\text{gnd}}}=\frac{1}{\sqrt{2}}(\ket{\Psi^+}-\ket{\Psi^-})\nonumber\\
=\frac{1}{\sqrt{2}}(\ket{\Psi^1_{\mathrm{even}}}\ket{\Psi^2_{\mathrm{odd}}}-\ket{\Psi^1_{\mathrm{odd}}}\ket{\Psi^2_{\mathrm{even}}})
\end{gather}

\begin{figure}[htbp]
\begin{center}
\includegraphics[scale=0.3]{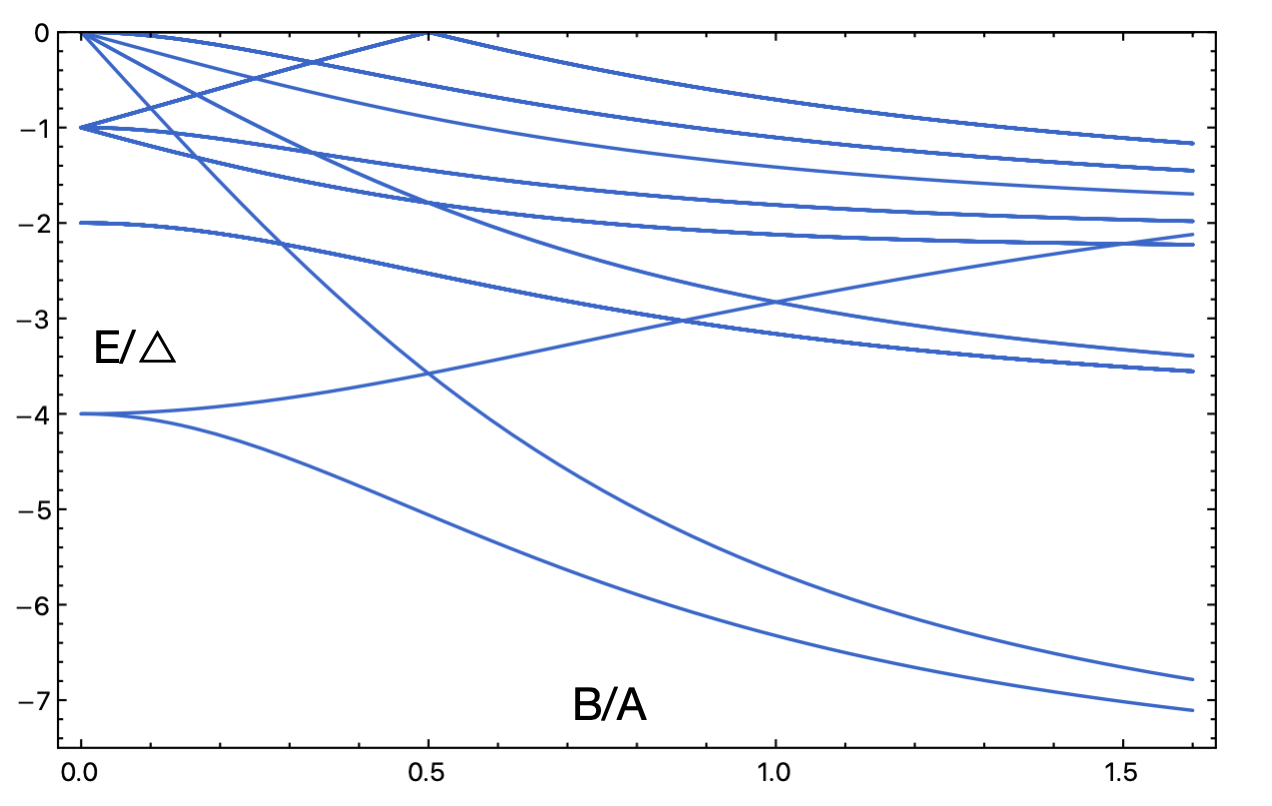}
\caption{Many-body energy levels in strong coupling limit of the single site quantum mechanics in 1+1d by
exact diagonalization (256$\times$256 matrix). X axis is $\frac{B}{A}$ and Y axis is $\frac{E}{\Delta}$, where $\Delta=\sqrt{A^2+B^2}$. When $\frac{B}{A}=0$, the ground state has two fold degeneracy as expected. Otherwise, the many-body gap never closes.}
\label{fig: band}
\end{center}
\end{figure}

Away from the perturbative limit, we can numerically check that the excitation gap never close as we tune the $B/A$ ratio, as shown in \figref{fig: band}. This implies that in the strong interaction limit, the two copies of the K\"ahler-Dirac fermion can indeed be gapped by the interaction without spontaneous symmetry breaking or topological ordering.

\section{Classification of Fermionic SPT with $\mathbb{Z}^P_2$ Symmetry}
In this section, our focus has been on the classification of fermionic SPT phases possessing inversion symmetry $\mathbb{Z}^P_2$. It is established that any two copies of $(d-1)+1$-dimensional massless Kähler-Dirac fermions, which serve as the boundary states of $d+1$-dimensional Kähler-Dirac fermions, can acquire a gap symmetrically. Consequently, our main objective is to determine the number of copies of root boundary fermions present in Kähler-Dirac fermions. The boundary state of a fermionic SPT phase in $d+1$ dimensions is characterized by the Hamiltonian $H=i\partial_i\gamma^i$, where $i=1,2,\ldots, d-1$, and $\gamma_i$ corresponds to the real Clifford algebra $\mathcal{C}l_{d-1,0}$. The dimension of the real irreducible representation of $\mathcal{C}l_{d-1,0}$ is denoted as $dim_{\mathbb{R}}\mathcal{C}l_{d-1,0}$. It is essential to note that Kähler-Dirac fermions comprise $2^d$ independent flavor fermions and have a real dimension of $dim_{\mathbb{R}}(\mathcal{H})=2dim_{\mathbb{C}}(\mathcal{H})=2\times2^d=2^{d+1}$. Thus, $(d-1)+1$-dimensional Kähler-Dirac fermions contain $\nu$ copies of the root boundary fermions, where 
\begin{equation}
    \nu=\frac{2^{d+1}}{dim_{\mathbb{R}}\mathcal{C}l_{d-1,0}}.
\end{equation}

Since $\nu$ copies of our root state are Kähler-Dirac fermions, and two copies of Kähler-Dirac fermions are trivial, the corresponding fermionic SPT phases are $\mathbb{Z}_{2\nu}$ classified.

Furthermore, the bulk state of SPT phases also undergoes SMG \cite{You:2014ho}. In the absence of interaction, the bulk is characterized by different topological numbers for distinct topological phases, and the bulk is gapless at the phase transition point. If the boundary state acquires mass symmetrically, the bulk does not undergo a phase transition either, and the original gapless phase transition point symmetrically acquires a mass gap as well. The parameter $\nu$ can be calculated from the bulk perspective as follows:

\begin{equation}
\nu=\frac{2^{d+2}}{dim_{\mathbb{R}}\mathcal{C}l_{d,1}}
\end{equation}
where $2^{d+2}$ is the dimension of bulk K\"ahler-Dirac fermions, and the root book fermions described by $i\partial_{\mu}\gamma^\mu\psi=0$ in Minkowski spacetime represent $\mathcal{C}l_{d,1}$.
\begin{table}[ht]
\centering
\caption{The classification of $\mathbb{Z}_2^P$ fermionic SPT phases. Where $\mathbb{R}$, $\mathbb{C}$, $\mathbb{H}$ denote real numbers, complex numbers and quaternions and the notation $\mathbb{F}(N)$ denotes the algebra of $N\times N$ matrices over $\mathbb{F}$. And $dim\mathbb{F}(N)=Ndim\mathbb{F}$. The result of classification in d+8 dimension comes from Bott periodicity.}
\begin{tabular}{ccccc}
\hline
d  & $\mathcal{C}l_{d-1,0}$&$\mathcal{C}l_{d,1}$& $\nu$ & classification \\
\hline
1 & $\mathbb{R}$ & $\mathbb{R}(2)$ & 4 & $\mathbb{Z}_8$ \\
2 & $\mathbb{R}\oplus \mathbb{R}$&$\mathbb{R}(2)\oplus \mathbb{R}(2)$ & 4 & $\mathbb{Z}_8$ \\
3 & $\mathbb{R}(2)$ &$\mathbb{R}(4)$  & 8 & $\mathbb{Z}_{16}$ \\
4 &  $\mathbb{C}(2)$ &$\mathbb{C}(4)$  &8 & $\mathbb{Z}_{16}$ \\
5 &  $\mathbb{H}(2)$&$\mathbb{H}(4)$  &8 & $\mathbb{Z}_{16}$ \\
6 & $\mathbb{H}(2)\oplus \mathbb{H}(2)$ &$\mathbb{H}(4)\oplus \mathbb{H}(4)$ &8 & $\mathbb{Z}_{16}$ \\
7 & $\mathbb{H}(4)$ & $\mathbb{H}(8)$ &16&$\mathbb{Z}_{32}$\\
8 &  $\mathbb{C}(8)$ &$\mathbb{C}(16)$ &32 & $\mathbb{Z}_{64}$ \\
\vdots&\vdots&\vdots&\vdots\\
d+8& $\mathcal{C}l_{d-1}\otimes\mathbb{R}(16)$& $\mathcal{C}l_{d,1}\otimes\mathbb{R}(16)$ &$16\nu_d$ & $\mathbb{Z}_{32\nu_d}$\\
\hline
\end{tabular}

\end{table}

\section{Conclusions}
In conclusion, we have explained the symmetric mass generation (SMG) of K\"ahler-Dirac fermions by extending the algebraic structure of the K\"ahler-Dirac field and defining a new operator, $\Tilde{K}$. Our study shows that K\"ahler-Dirac fermions correspond to bosonic symmetry protected topological (SPT) phases, specifically $\mathbb{Z}_2$ SPT phases protected by $\mathbb{Z}_2^P$ symmetry. Constructing an $O(D+2)$ nonlinear sigma model (NL$\sigma$M) with a topological term, we have unveiled connections between interacting K\"ahler-Dirac fermions and condensed matter physics, as well as applications in classifying interacting fermionic SPT phases.

Moreover, our approach to constructing bosonic SPT phases has proven to be applicable to lattice models. We demonstrate that the boundary of these fermions consists of massless K\"ahler-Dirac fermions. Utilizing this relationship, we have explored the bulk boundary correspondence and discussed the classification of $\mathbb{Z}_2$ fermionic SPT. Our findings reveal that only an even number of K\"ahler-Dirac fermion copies can symmetrically acquire a mass gap, regardless of the spacetime dimensions in which they are defined. To be specific, eight copies of massless Dirac fermions can symmetrically gap in (3+1)d space and time. This work broadens our understanding of symmetric mass generation, providing valuable insights into the topological properties of interacting fermionic systems.

\begin{acknowledgments}
This research is supported by the NSF Grant DMR-2238360.
\end{acknowledgments}

\bibliography{main}

\providecommand{\noopsort}[1]{}\providecommand{\singleletter}[1]{#1}%
\begin{thebibliography}{90}%
\makeatletter
\providecommand \@ifxundefined [1]{%
 \@ifx{#1\undefined}
}%
\providecommand \@ifnum [1]{%
 \ifnum #1\expandafter \@firstoftwo
 \else \expandafter \@secondoftwo
 \fi
}%
\providecommand \@ifx [1]{%
 \ifx #1\expandafter \@firstoftwo
 \else \expandafter \@secondoftwo
 \fi
}%
\providecommand \natexlab [1]{#1}%
\providecommand \enquote  [1]{``#1''}%
\providecommand \bibnamefont  [1]{#1}%
\providecommand \bibfnamefont [1]{#1}%
\providecommand \citenamefont [1]{#1}%
\providecommand \href@noop [0]{\@secondoftwo}%
\providecommand \href [0]{\begingroup \@sanitize@url \@href}%
\providecommand \@href[1]{\@@startlink{#1}\@@href}%
\providecommand \@@href[1]{\endgroup#1\@@endlink}%
\providecommand \@sanitize@url [0]{\catcode `\\12\catcode `\$12\catcode
  `\&12\catcode `\#12\catcode `\^12\catcode `\_12\catcode `\%12\relax}%
\providecommand \@@startlink[1]{}%
\providecommand \@@endlink[0]{}%
\providecommand \url  [0]{\begingroup\@sanitize@url \@url }%
\providecommand \@url [1]{\endgroup\@href {#1}{\urlprefix }}%
\providecommand \urlprefix  [0]{URL }%
\providecommand \Eprint [0]{\href }%
\providecommand \doibase [0]{https://doi.org/}%
\providecommand \selectlanguage [0]{\@gobble}%
\providecommand \bibinfo  [0]{\@secondoftwo}%
\providecommand \bibfield  [0]{\@secondoftwo}%
\providecommand \translation [1]{[#1]}%
\providecommand \BibitemOpen [0]{}%
\providecommand \bibitemStop [0]{}%
\providecommand \bibitemNoStop [0]{.\EOS\space}%
\providecommand \EOS [0]{\spacefactor3000\relax}%
\providecommand \BibitemShut  [1]{\csname bibitem#1\endcsname}%
\let\auto@bib@innerbib\@empty
\bibitem [{\citenamefont {{Fidkowski}}\ and\ \citenamefont
  {{Kitaev}}(2010)}]{Fidkowski:2010bf}%
  \BibitemOpen
  \bibfield  {author} {\bibinfo {author} {\bibfnamefont {L.}~\bibnamefont
  {{Fidkowski}}}\ and\ \bibinfo {author} {\bibfnamefont {A.}~\bibnamefont
  {{Kitaev}}},\ }\bibfield  {title} {\bibinfo {title} {{Effects of interactions
  on the topological classification of free fermion systems}},\ }\href
  {https://doi.org/10.1103/PhysRevB.81.134509} {\bibfield  {journal} {\bibinfo
  {journal} {\prb}\ }\textbf {\bibinfo {volume} {81}},\ \bibinfo {eid} {134509}
  (\bibinfo {year} {2010})},\ \Eprint {https://arxiv.org/abs/0904.2197}
  {arXiv:0904.2197 [cond-mat.str-el]} \BibitemShut {NoStop}%
\bibitem [{\citenamefont {{Fidkowski}}\ and\ \citenamefont
  {{Kitaev}}(2011)}]{Fidkowski:2011dd}%
  \BibitemOpen
  \bibfield  {author} {\bibinfo {author} {\bibfnamefont {L.}~\bibnamefont
  {{Fidkowski}}}\ and\ \bibinfo {author} {\bibfnamefont {A.}~\bibnamefont
  {{Kitaev}}},\ }\bibfield  {title} {\bibinfo {title} {{Topological phases of
  fermions in one dimension}},\ }\href
  {https://doi.org/10.1103/PhysRevB.83.075103} {\bibfield  {journal} {\bibinfo
  {journal} {\prb}\ }\textbf {\bibinfo {volume} {83}},\ \bibinfo {eid} {075103}
  (\bibinfo {year} {2011})},\ \Eprint {https://arxiv.org/abs/1008.4138}
  {arXiv:1008.4138 [cond-mat.str-el]} \BibitemShut {NoStop}%
\bibitem [{\citenamefont {{Wang}}\ and\ \citenamefont
  {{Wen}}(2013)}]{Wang2013Non-Perturbative}%
  \BibitemOpen
  \bibfield  {author} {\bibinfo {author} {\bibfnamefont {J.}~\bibnamefont
  {{Wang}}}\ and\ \bibinfo {author} {\bibfnamefont {X.-G.}\ \bibnamefont
  {{Wen}}},\ }\bibfield  {title} {\bibinfo {title} {{Non-Perturbative
  Regularization of 1+1D Anomaly-Free Chiral Fermions and Bosons: On the
  equivalence of anomaly matching conditions and boundary gapping rules}},\
  }\href@noop {} {\bibfield  {journal} {\bibinfo  {journal} {arXiv e-prints}\
  ,\ \bibinfo {eid} {arXiv:1307.7480}} (\bibinfo {year} {2013})},\ \Eprint
  {https://arxiv.org/abs/1307.7480} {arXiv:1307.7480 [hep-lat]} \BibitemShut
  {NoStop}%
\bibitem [{\citenamefont {{Slagle}}\ \emph {et~al.}(2015)\citenamefont
  {{Slagle}}, \citenamefont {{You}},\ and\ \citenamefont
  {{Xu}}}]{Slagle:2015lo}%
  \BibitemOpen
  \bibfield  {author} {\bibinfo {author} {\bibfnamefont {K.}~\bibnamefont
  {{Slagle}}}, \bibinfo {author} {\bibfnamefont {Y.-Z.}\ \bibnamefont
  {{You}}},\ and\ \bibinfo {author} {\bibfnamefont {C.}~\bibnamefont {{Xu}}},\
  }\bibfield  {title} {\bibinfo {title} {{Exotic quantum phase transitions of
  strongly interacting topological insulators}},\ }\href
  {https://doi.org/10.1103/PhysRevB.91.115121} {\bibfield  {journal} {\bibinfo
  {journal} {\prb}\ }\textbf {\bibinfo {volume} {91}},\ \bibinfo {eid} {115121}
  (\bibinfo {year} {2015})},\ \Eprint {https://arxiv.org/abs/1409.7401}
  {arXiv:1409.7401 [cond-mat.str-el]} \BibitemShut {NoStop}%
\bibitem [{\citenamefont {{Ayyar}}\ and\ \citenamefont
  {{Chandrasekharan}}(2015)}]{Ayyar2015Massive}%
  \BibitemOpen
  \bibfield  {author} {\bibinfo {author} {\bibfnamefont {V.}~\bibnamefont
  {{Ayyar}}}\ and\ \bibinfo {author} {\bibfnamefont {S.}~\bibnamefont
  {{Chandrasekharan}}},\ }\bibfield  {title} {\bibinfo {title} {{Massive
  fermions without fermion bilinear condensates}},\ }\href
  {https://doi.org/10.1103/PhysRevD.91.065035} {\bibfield  {journal} {\bibinfo
  {journal} {\prd}\ }\textbf {\bibinfo {volume} {91}},\ \bibinfo {eid} {065035}
  (\bibinfo {year} {2015})},\ \Eprint {https://arxiv.org/abs/1410.6474}
  {arXiv:1410.6474 [hep-lat]} \BibitemShut {NoStop}%
\bibitem [{\citenamefont {{Catterall}}(2016)}]{Catterall:2016sw}%
  \BibitemOpen
  \bibfield  {author} {\bibinfo {author} {\bibfnamefont {S.}~\bibnamefont
  {{Catterall}}},\ }\bibfield  {title} {\bibinfo {title} {{Fermion mass without
  symmetry breaking}},\ }\href {https://doi.org/10.1007/JHEP01(2016)121}
  {\bibfield  {journal} {\bibinfo  {journal} {Journal of High Energy Physics}\
  }\textbf {\bibinfo {volume} {1}},\ \bibinfo {eid} {121} (\bibinfo {year}
  {2016})},\ \Eprint {https://arxiv.org/abs/1510.04153} {arXiv:1510.04153
  [hep-lat]} \BibitemShut {NoStop}%
\bibitem [{\citenamefont {{Tong}}(2021)}]{Tong2021Comments}%
  \BibitemOpen
  \bibfield  {author} {\bibinfo {author} {\bibfnamefont {D.}~\bibnamefont
  {{Tong}}},\ }\bibfield  {title} {\bibinfo {title} {{Comments on Symmetric
  Mass Generation in 2d and 4d}},\ }\href@noop {} {\bibfield  {journal}
  {\bibinfo  {journal} {arXiv e-prints}\ ,\ \bibinfo {eid} {arXiv:2104.03997}}
  (\bibinfo {year} {2021})},\ \Eprint {https://arxiv.org/abs/2104.03997}
  {arXiv:2104.03997 [hep-th]} \BibitemShut {NoStop}%
\bibitem [{\citenamefont {{Wang}}\ and\ \citenamefont
  {{You}}(2022)}]{Wang2204.14271}%
  \BibitemOpen
  \bibfield  {author} {\bibinfo {author} {\bibfnamefont {J.}~\bibnamefont
  {{Wang}}}\ and\ \bibinfo {author} {\bibfnamefont {Y.-Z.}\ \bibnamefont
  {{You}}},\ }\bibfield  {title} {\bibinfo {title} {{Symmetric Mass
  Generation}},\ }\href {https://doi.org/10.3390/sym14071475} {\bibfield
  {journal} {\bibinfo  {journal} {Symmetry}\ }\textbf {\bibinfo {volume}
  {14}},\ \bibinfo {pages} {1475} (\bibinfo {year} {2022})},\ \Eprint
  {https://arxiv.org/abs/2204.14271} {arXiv:2204.14271 [cond-mat.str-el]}
  \BibitemShut {NoStop}%
\bibitem [{\citenamefont {{Ryu}}\ \emph {et~al.}(2012)\citenamefont {{Ryu}},
  \citenamefont {{Moore}},\ and\ \citenamefont
  {{Ludwig}}}]{Ryu2012Electromagnetic}%
  \BibitemOpen
  \bibfield  {author} {\bibinfo {author} {\bibfnamefont {S.}~\bibnamefont
  {{Ryu}}}, \bibinfo {author} {\bibfnamefont {J.~E.}\ \bibnamefont {{Moore}}},\
  and\ \bibinfo {author} {\bibfnamefont {A.~W.~W.}\ \bibnamefont {{Ludwig}}},\
  }\bibfield  {title} {\bibinfo {title} {{Electromagnetic and gravitational
  responses and anomalies in topological insulators and superconductors}},\
  }\href {https://doi.org/10.1103/PhysRevB.85.045104} {\bibfield  {journal}
  {\bibinfo  {journal} {\prb}\ }\textbf {\bibinfo {volume} {85}},\ \bibinfo
  {eid} {045104} (\bibinfo {year} {2012})},\ \Eprint
  {https://arxiv.org/abs/1010.0936} {arXiv:1010.0936 [cond-mat.str-el]}
  \BibitemShut {NoStop}%
\bibitem [{\citenamefont {{Kapustin}}\ and\ \citenamefont
  {{Thorngren}}(2014)}]{Kapustin2014Anomalies}%
  \BibitemOpen
  \bibfield  {author} {\bibinfo {author} {\bibfnamefont {A.}~\bibnamefont
  {{Kapustin}}}\ and\ \bibinfo {author} {\bibfnamefont {R.}~\bibnamefont
  {{Thorngren}}},\ }\bibfield  {title} {\bibinfo {title} {{Anomalies of
  discrete symmetries in various dimensions and group cohomology}},\
  }\href@noop {} {\bibfield  {journal} {\bibinfo  {journal} {arXiv e-prints}\
  ,\ \bibinfo {eid} {arXiv:1404.3230}} (\bibinfo {year} {2014})},\ \Eprint
  {https://arxiv.org/abs/1404.3230} {arXiv:1404.3230 [hep-th]} \BibitemShut
  {NoStop}%
\bibitem [{\citenamefont {{Tanizaki}}(2018)}]{Tanizaki2018Anomaly}%
  \BibitemOpen
  \bibfield  {author} {\bibinfo {author} {\bibfnamefont {Y.}~\bibnamefont
  {{Tanizaki}}},\ }\bibfield  {title} {\bibinfo {title} {{Anomaly constraint on
  massless QCD and the role of Skyrmions in chiral symmetry breaking}},\ }\href
  {https://doi.org/10.1007/JHEP08(2018)171} {\bibfield  {journal} {\bibinfo
  {journal} {Journal of High Energy Physics}\ }\textbf {\bibinfo {volume}
  {2018}},\ \bibinfo {eid} {171} (\bibinfo {year} {2018})},\ \Eprint
  {https://arxiv.org/abs/1807.07666} {arXiv:1807.07666 [hep-th]} \BibitemShut
  {NoStop}%
\bibitem [{\citenamefont {{Tachikawa}}\ and\ \citenamefont
  {{Yonekura}}(2019)}]{Tachikawa1805.02772}%
  \BibitemOpen
  \bibfield  {author} {\bibinfo {author} {\bibfnamefont {Y.}~\bibnamefont
  {{Tachikawa}}}\ and\ \bibinfo {author} {\bibfnamefont {K.}~\bibnamefont
  {{Yonekura}}},\ }\bibfield  {title} {\bibinfo {title} {{Why are fractional
  charges of orientifolds compatible with Dirac quantization?}},\ }\href
  {https://doi.org/10.21468/SciPostPhys.7.5.058} {\bibfield  {journal}
  {\bibinfo  {journal} {SciPost Physics}\ }\textbf {\bibinfo {volume} {7}},\
  \bibinfo {eid} {058} (\bibinfo {year} {2019})},\ \Eprint
  {https://arxiv.org/abs/1805.02772} {arXiv:1805.02772 [hep-th]} \BibitemShut
  {NoStop}%
\bibitem [{\citenamefont {{Yamaguchi}}(2019)}]{Yamaguchi2019t-Hooft}%
  \BibitemOpen
  \bibfield  {author} {\bibinfo {author} {\bibfnamefont {S.}~\bibnamefont
  {{Yamaguchi}}},\ }\bibfield  {title} {\bibinfo {title} {{'t Hooft anomaly
  matching condition and chiral symmetry breaking without bilinear
  condensate}},\ }\href {https://doi.org/10.1007/JHEP01(2019)014} {\bibfield
  {journal} {\bibinfo  {journal} {Journal of High Energy Physics}\ }\textbf
  {\bibinfo {volume} {2019}},\ \bibinfo {eid} {14} (\bibinfo {year} {2019})},\
  \Eprint {https://arxiv.org/abs/1811.09390} {arXiv:1811.09390 [hep-th]}
  \BibitemShut {NoStop}%
\bibitem [{\citenamefont {{Ayyar}}\ and\ \citenamefont
  {{Chandrasekharan}}(2016{\natexlab{a}})}]{Ayyar:2016fi}%
  \BibitemOpen
  \bibfield  {author} {\bibinfo {author} {\bibfnamefont {V.}~\bibnamefont
  {{Ayyar}}}\ and\ \bibinfo {author} {\bibfnamefont {S.}~\bibnamefont
  {{Chandrasekharan}}},\ }\bibfield  {title} {\bibinfo {title} {{Origin of
  fermion masses without spontaneous symmetry breaking}},\ }\href
  {https://doi.org/10.1103/PhysRevD.93.081701} {\bibfield  {journal} {\bibinfo
  {journal} {\prd}\ }\textbf {\bibinfo {volume} {93}},\ \bibinfo {eid} {081701}
  (\bibinfo {year} {2016}{\natexlab{a}})},\ \Eprint
  {https://arxiv.org/abs/1511.09071} {arXiv:1511.09071 [hep-lat]} \BibitemShut
  {NoStop}%
\bibitem [{\citenamefont {{Catterall}}\ and\ \citenamefont
  {{Schaich}}(2016)}]{Catterall:2016nh}%
  \BibitemOpen
  \bibfield  {author} {\bibinfo {author} {\bibfnamefont {S.}~\bibnamefont
  {{Catterall}}}\ and\ \bibinfo {author} {\bibfnamefont {D.}~\bibnamefont
  {{Schaich}}},\ }\bibfield  {title} {\bibinfo {title} {{Novel phases in
  strongly coupled four-fermion theories}},\ }\href@noop {} {\bibfield
  {journal} {\bibinfo  {journal} {ArXiv e-prints}\ } (\bibinfo {year}
  {2016})},\ \Eprint {https://arxiv.org/abs/1609.08541} {arXiv:1609.08541
  [hep-lat]} \BibitemShut {NoStop}%
\bibitem [{\citenamefont {{Ayyar}}\ and\ \citenamefont
  {{Chandrasekharan}}(2016{\natexlab{b}})}]{Ayyar:2016tg}%
  \BibitemOpen
  \bibfield  {author} {\bibinfo {author} {\bibfnamefont {V.}~\bibnamefont
  {{Ayyar}}}\ and\ \bibinfo {author} {\bibfnamefont {S.}~\bibnamefont
  {{Chandrasekharan}}},\ }\bibfield  {title} {\bibinfo {title} {{Fermion masses
  through four-fermion condensates}},\ }\href
  {https://doi.org/10.1007/JHEP10(2016)058} {\bibfield  {journal} {\bibinfo
  {journal} {Journal of High Energy Physics}\ }\textbf {\bibinfo {volume}
  {10}},\ \bibinfo {eid} {58} (\bibinfo {year} {2016}{\natexlab{b}})},\ \Eprint
  {https://arxiv.org/abs/1606.06312} {arXiv:1606.06312 [hep-lat]} \BibitemShut
  {NoStop}%
\bibitem [{\citenamefont {{Witten}}(2016)}]{Witten:2016yb}%
  \BibitemOpen
  \bibfield  {author} {\bibinfo {author} {\bibfnamefont {E.}~\bibnamefont
  {{Witten}}},\ }\bibfield  {title} {\bibinfo {title} {{The ``parity'' anomaly
  on an unorientable manifold}},\ }\href
  {https://doi.org/10.1103/PhysRevB.94.195150} {\bibfield  {journal} {\bibinfo
  {journal} {\prb}\ }\textbf {\bibinfo {volume} {94}},\ \bibinfo {eid} {195150}
  (\bibinfo {year} {2016})},\ \Eprint {https://arxiv.org/abs/1605.02391}
  {arXiv:1605.02391 [hep-th]} \BibitemShut {NoStop}%
\bibitem [{\citenamefont {{Ayyar}}(2016)}]{Ayyar:2016ph}%
  \BibitemOpen
  \bibfield  {author} {\bibinfo {author} {\bibfnamefont {V.}~\bibnamefont
  {{Ayyar}}},\ }\bibfield  {title} {\bibinfo {title} {{Search for a continuum
  limit of the PMS phase}},\ }\href@noop {} {\bibfield  {journal} {\bibinfo
  {journal} {ArXiv e-prints}\ } (\bibinfo {year} {2016})},\ \Eprint
  {https://arxiv.org/abs/1611.00280} {arXiv:1611.00280 [hep-lat]} \BibitemShut
  {NoStop}%
\bibitem [{\citenamefont {{He}}\ \emph {et~al.}(2016)\citenamefont {{He}},
  \citenamefont {{Wu}}, \citenamefont {{You}}, \citenamefont {{Xu}},
  \citenamefont {{Meng}},\ and\ \citenamefont {{Lu}}}]{He:2016qy}%
  \BibitemOpen
  \bibfield  {author} {\bibinfo {author} {\bibfnamefont {Y.-Y.}\ \bibnamefont
  {{He}}}, \bibinfo {author} {\bibfnamefont {H.-Q.}\ \bibnamefont {{Wu}}},
  \bibinfo {author} {\bibfnamefont {Y.-Z.}\ \bibnamefont {{You}}}, \bibinfo
  {author} {\bibfnamefont {C.}~\bibnamefont {{Xu}}}, \bibinfo {author}
  {\bibfnamefont {Z.~Y.}\ \bibnamefont {{Meng}}},\ and\ \bibinfo {author}
  {\bibfnamefont {Z.-Y.}\ \bibnamefont {{Lu}}},\ }\bibfield  {title} {\bibinfo
  {title} {{Quantum critical point of Dirac fermion mass generation without
  spontaneous symmetry breaking}},\ }\href
  {https://doi.org/10.1103/PhysRevB.94.241111} {\bibfield  {journal} {\bibinfo
  {journal} {\prb}\ }\textbf {\bibinfo {volume} {94}},\ \bibinfo {eid} {241111}
  (\bibinfo {year} {2016})},\ \Eprint {https://arxiv.org/abs/1603.08376}
  {arXiv:1603.08376 [cond-mat.str-el]} \BibitemShut {NoStop}%
\bibitem [{\citenamefont {{Ayyar}}\ and\ \citenamefont
  {{Chandrasekharan}}(2017)}]{Ayyar2017Generating}%
  \BibitemOpen
  \bibfield  {author} {\bibinfo {author} {\bibfnamefont {V.}~\bibnamefont
  {{Ayyar}}}\ and\ \bibinfo {author} {\bibfnamefont {S.}~\bibnamefont
  {{Chandrasekharan}}},\ }\bibfield  {title} {\bibinfo {title} {{Generating a
  nonperturbative mass gap using Feynman diagrams in an asymptotically free
  theory}},\ }\href {https://doi.org/10.1103/PhysRevD.96.114506} {\bibfield
  {journal} {\bibinfo  {journal} {\prd}\ }\textbf {\bibinfo {volume} {96}},\
  \bibinfo {eid} {114506} (\bibinfo {year} {2017})},\ \Eprint
  {https://arxiv.org/abs/1709.06048} {arXiv:1709.06048 [hep-lat]} \BibitemShut
  {NoStop}%
\bibitem [{\citenamefont {{You}}\ \emph
  {et~al.}(2018{\natexlab{a}})\citenamefont {{You}}, \citenamefont {{He}},
  \citenamefont {{Xu}},\ and\ \citenamefont {{Vishwanath}}}]{You2018Symmetric}%
  \BibitemOpen
  \bibfield  {author} {\bibinfo {author} {\bibfnamefont {Y.-Z.}\ \bibnamefont
  {{You}}}, \bibinfo {author} {\bibfnamefont {Y.-C.}\ \bibnamefont {{He}}},
  \bibinfo {author} {\bibfnamefont {C.}~\bibnamefont {{Xu}}},\ and\ \bibinfo
  {author} {\bibfnamefont {A.}~\bibnamefont {{Vishwanath}}},\ }\bibfield
  {title} {\bibinfo {title} {{Symmetric Fermion Mass Generation as Deconfined
  Quantum Criticality}},\ }\href {https://doi.org/10.1103/PhysRevX.8.011026}
  {\bibfield  {journal} {\bibinfo  {journal} {Physical Review X}\ }\textbf
  {\bibinfo {volume} {8}},\ \bibinfo {eid} {011026} (\bibinfo {year}
  {2018}{\natexlab{a}})},\ \Eprint {https://arxiv.org/abs/1705.09313}
  {arXiv:1705.09313 [cond-mat.str-el]} \BibitemShut {NoStop}%
\bibitem [{\citenamefont {{You}}\ \emph
  {et~al.}(2018{\natexlab{b}})\citenamefont {{You}}, \citenamefont {{He}},
  \citenamefont {{Vishwanath}},\ and\ \citenamefont {{Xu}}}]{You2018From}%
  \BibitemOpen
  \bibfield  {author} {\bibinfo {author} {\bibfnamefont {Y.-Z.}\ \bibnamefont
  {{You}}}, \bibinfo {author} {\bibfnamefont {Y.-C.}\ \bibnamefont {{He}}},
  \bibinfo {author} {\bibfnamefont {A.}~\bibnamefont {{Vishwanath}}},\ and\
  \bibinfo {author} {\bibfnamefont {C.}~\bibnamefont {{Xu}}},\ }\bibfield
  {title} {\bibinfo {title} {{From bosonic topological transition to symmetric
  fermion mass generation}},\ }\href
  {https://doi.org/10.1103/PhysRevB.97.125112} {\bibfield  {journal} {\bibinfo
  {journal} {\prb}\ }\textbf {\bibinfo {volume} {97}},\ \bibinfo {eid} {125112}
  (\bibinfo {year} {2018}{\natexlab{b}})},\ \Eprint
  {https://arxiv.org/abs/1711.00863} {arXiv:1711.00863 [cond-mat.str-el]}
  \BibitemShut {NoStop}%
\bibitem [{\citenamefont {{Schaich}}\ and\ \citenamefont
  {{Catterall}}(2018)}]{Schaich2018Phases}%
  \BibitemOpen
  \bibfield  {author} {\bibinfo {author} {\bibfnamefont {D.}~\bibnamefont
  {{Schaich}}}\ and\ \bibinfo {author} {\bibfnamefont {S.}~\bibnamefont
  {{Catterall}}},\ }\bibfield  {title} {\bibinfo {title} {{Phases of a strongly
  coupled four-fermion theory}},\ }in\ \href
  {https://doi.org/10.1051/epjconf/201817503004} {\emph {\bibinfo {booktitle}
  {European Physical Journal Web of Conferences}}},\ \bibinfo {series}
  {European Physical Journal Web of Conferences}, Vol.\ \bibinfo {volume}
  {175}\ (\bibinfo {year} {2018})\ p.\ \bibinfo {pages} {03004},\ \Eprint
  {https://arxiv.org/abs/1710.08137} {arXiv:1710.08137 [hep-lat]} \BibitemShut
  {NoStop}%
\bibitem [{\citenamefont {{Catterall}}\ and\ \citenamefont
  {{Butt}}(2018)}]{Catterall2018Topology}%
  \BibitemOpen
  \bibfield  {author} {\bibinfo {author} {\bibfnamefont {S.}~\bibnamefont
  {{Catterall}}}\ and\ \bibinfo {author} {\bibfnamefont {N.}~\bibnamefont
  {{Butt}}},\ }\bibfield  {title} {\bibinfo {title} {{Topology and strong four
  fermion interactions in four dimensions}},\ }\href
  {https://doi.org/10.1103/PhysRevD.97.094502} {\bibfield  {journal} {\bibinfo
  {journal} {\prd}\ }\textbf {\bibinfo {volume} {97}},\ \bibinfo {eid} {094502}
  (\bibinfo {year} {2018})},\ \Eprint {https://arxiv.org/abs/1708.06715}
  {arXiv:1708.06715 [hep-lat]} \BibitemShut {NoStop}%
\bibitem [{\citenamefont {{Butt}}\ and\ \citenamefont
  {{Catterall}}(2018)}]{Butt2018Four}%
  \BibitemOpen
  \bibfield  {author} {\bibinfo {author} {\bibfnamefont {N.}~\bibnamefont
  {{Butt}}}\ and\ \bibinfo {author} {\bibfnamefont {S.}~\bibnamefont
  {{Catterall}}},\ }\bibfield  {title} {\bibinfo {title} {{Four fermion
  condensates in SU(2) Yang-Mills-Higgs theory on a lattice}},\ }in\ \href@noop
  {} {\emph {\bibinfo {booktitle} {The 36th Annual International Symposium on
  Lattice Field Theory. 22-28 July}}}\ (\bibinfo {year} {2018})\ p.\ \bibinfo
  {pages} {294},\ \Eprint {https://arxiv.org/abs/1811.01015} {arXiv:1811.01015
  [hep-lat]} \BibitemShut {NoStop}%
\bibitem [{\citenamefont {{Butt}}\ \emph {et~al.}(2018)\citenamefont {{Butt}},
  \citenamefont {{Catterall}},\ and\ \citenamefont
  {{Schaich}}}]{Butt2018SO4-invariant}%
  \BibitemOpen
  \bibfield  {author} {\bibinfo {author} {\bibfnamefont {N.}~\bibnamefont
  {{Butt}}}, \bibinfo {author} {\bibfnamefont {S.}~\bibnamefont
  {{Catterall}}},\ and\ \bibinfo {author} {\bibfnamefont {D.}~\bibnamefont
  {{Schaich}}},\ }\bibfield  {title} {\bibinfo {title} {{SO(4) invariant
  Higgs-Yukawa model with reduced staggered fermions}},\ }\href
  {https://doi.org/10.1103/PhysRevD.98.114514} {\bibfield  {journal} {\bibinfo
  {journal} {\prd}\ }\textbf {\bibinfo {volume} {98}},\ \bibinfo {eid} {114514}
  (\bibinfo {year} {2018})},\ \Eprint {https://arxiv.org/abs/1810.06117}
  {arXiv:1810.06117 [hep-lat]} \BibitemShut {NoStop}%
\bibitem [{\citenamefont {{Catterall}}\ \emph {et~al.}(2020)\citenamefont
  {{Catterall}}, \citenamefont {{Butt}},\ and\ \citenamefont
  {{Schaich}}}]{Catterall2020Exotic}%
  \BibitemOpen
  \bibfield  {author} {\bibinfo {author} {\bibfnamefont {S.}~\bibnamefont
  {{Catterall}}}, \bibinfo {author} {\bibfnamefont {N.}~\bibnamefont
  {{Butt}}},\ and\ \bibinfo {author} {\bibfnamefont {D.}~\bibnamefont
  {{Schaich}}},\ }\bibfield  {title} {\bibinfo {title} {{Exotic Phases of a
  Higgs-Yukawa Model with Reduced Staggered Fermions}},\ }\href@noop {}
  {\bibfield  {journal} {\bibinfo  {journal} {arXiv e-prints}\ ,\ \bibinfo
  {eid} {arXiv:2002.00034}} (\bibinfo {year} {2020})},\ \Eprint
  {https://arxiv.org/abs/2002.00034} {arXiv:2002.00034 [hep-lat]} \BibitemShut
  {NoStop}%
\bibitem [{\citenamefont {{Xu}}\ and\ \citenamefont
  {{Xu}}(2021)}]{Xu2021Greens}%
  \BibitemOpen
  \bibfield  {author} {\bibinfo {author} {\bibfnamefont {Y.}~\bibnamefont
  {{Xu}}}\ and\ \bibinfo {author} {\bibfnamefont {C.}~\bibnamefont {{Xu}}},\
  }\bibfield  {title} {\bibinfo {title} {{Green's function Zero and Symmetric
  Mass Generation}},\ }\href@noop {} {\bibfield  {journal} {\bibinfo  {journal}
  {arXiv e-prints}\ ,\ \bibinfo {eid} {arXiv:2103.15865}} (\bibinfo {year}
  {2021})},\ \Eprint {https://arxiv.org/abs/2103.15865} {arXiv:2103.15865
  [cond-mat.str-el]} \BibitemShut {NoStop}%
\bibitem [{\citenamefont {{Catterall}}(2021)}]{Catterall2021Chiral}%
  \BibitemOpen
  \bibfield  {author} {\bibinfo {author} {\bibfnamefont {S.}~\bibnamefont
  {{Catterall}}},\ }\bibfield  {title} {\bibinfo {title} {{Chiral lattice
  fermions from staggered fields}},\ }\href
  {https://doi.org/10.1103/PhysRevD.104.014503} {\bibfield  {journal} {\bibinfo
   {journal} {\prd}\ }\textbf {\bibinfo {volume} {104}},\ \bibinfo {eid}
  {014503} (\bibinfo {year} {2021})},\ \Eprint
  {https://arxiv.org/abs/2010.02290} {arXiv:2010.02290 [hep-lat]} \BibitemShut
  {NoStop}%
\bibitem [{\citenamefont {{Butt}}\ \emph
  {et~al.}(2021{\natexlab{a}})\citenamefont {{Butt}}, \citenamefont
  {{Catterall}},\ and\ \citenamefont {{Toga}}}]{Butt2021Symmetric}%
  \BibitemOpen
  \bibfield  {author} {\bibinfo {author} {\bibfnamefont {N.}~\bibnamefont
  {{Butt}}}, \bibinfo {author} {\bibfnamefont {S.}~\bibnamefont
  {{Catterall}}},\ and\ \bibinfo {author} {\bibfnamefont {G.~C.}\ \bibnamefont
  {{Toga}}},\ }\bibfield  {title} {\bibinfo {title} {{Symmetric Mass Generation
  in Lattice Gauge Theory}},\ }\href@noop {} {\bibfield  {journal} {\bibinfo
  {journal} {arXiv e-prints}\ ,\ \bibinfo {eid} {arXiv:2111.01001}} (\bibinfo
  {year} {2021}{\natexlab{a}})},\ \Eprint {https://arxiv.org/abs/2111.01001}
  {arXiv:2111.01001 [hep-lat]} \BibitemShut {NoStop}%
\bibitem [{\citenamefont {{Lu}}\ \emph {et~al.}(2022)\citenamefont {{Lu}},
  \citenamefont {{Zeng}}, \citenamefont {{Wang}},\ and\ \citenamefont
  {{You}}}]{Lu2210.16304}%
  \BibitemOpen
  \bibfield  {author} {\bibinfo {author} {\bibfnamefont {D.-C.}\ \bibnamefont
  {{Lu}}}, \bibinfo {author} {\bibfnamefont {M.}~\bibnamefont {{Zeng}}},
  \bibinfo {author} {\bibfnamefont {J.}~\bibnamefont {{Wang}}},\ and\ \bibinfo
  {author} {\bibfnamefont {Y.-Z.}\ \bibnamefont {{You}}},\ }\bibfield  {title}
  {\bibinfo {title} {{Fermi Surface Symmetric Mass Generation}},\ }\href
  {https://doi.org/10.48550/arXiv.2210.16304} {\bibfield  {journal} {\bibinfo
  {journal} {arXiv e-prints}\ ,\ \bibinfo {eid} {arXiv:2210.16304}} (\bibinfo
  {year} {2022})},\ \Eprint {https://arxiv.org/abs/2210.16304}
  {arXiv:2210.16304 [cond-mat.str-el]} \BibitemShut {NoStop}%
\bibitem [{\citenamefont {{Gu}}\ and\ \citenamefont
  {{Wen}}(2012)}]{Gu2012Symmetry-protected}%
  \BibitemOpen
  \bibfield  {author} {\bibinfo {author} {\bibfnamefont {Z.-C.}\ \bibnamefont
  {{Gu}}}\ and\ \bibinfo {author} {\bibfnamefont {X.-G.}\ \bibnamefont
  {{Wen}}},\ }\bibfield  {title} {\bibinfo {title} {{Symmetry-protected
  topological orders for interacting fermions -- Fermionic topological
  nonlinear $\sigma$ models and a special group supercohomology theory}},\
  }\href@noop {} {\bibfield  {journal} {\bibinfo  {journal} {arXiv e-prints}\
  ,\ \bibinfo {eid} {arXiv:1201.2648}} (\bibinfo {year} {2012})},\ \Eprint
  {https://arxiv.org/abs/1201.2648} {arXiv:1201.2648 [cond-mat.str-el]}
  \BibitemShut {NoStop}%
\bibitem [{\citenamefont {{Cheng}}\ \emph {et~al.}(2015)\citenamefont
  {{Cheng}}, \citenamefont {{Bi}}, \citenamefont {{You}},\ and\ \citenamefont
  {{Gu}}}]{Cheng2015Classification}%
  \BibitemOpen
  \bibfield  {author} {\bibinfo {author} {\bibfnamefont {M.}~\bibnamefont
  {{Cheng}}}, \bibinfo {author} {\bibfnamefont {Z.}~\bibnamefont {{Bi}}},
  \bibinfo {author} {\bibfnamefont {Y.-Z.}\ \bibnamefont {{You}}},\ and\
  \bibinfo {author} {\bibfnamefont {Z.-C.}\ \bibnamefont {{Gu}}},\ }\bibfield
  {title} {\bibinfo {title} {{Classification of Symmetry-Protected Phases for
  Interacting Fermions in Two Dimensions}},\ }\href@noop {} {\bibfield
  {journal} {\bibinfo  {journal} {arXiv e-prints}\ ,\ \bibinfo {eid}
  {arXiv:1501.01313}} (\bibinfo {year} {2015})},\ \Eprint
  {https://arxiv.org/abs/1501.01313} {arXiv:1501.01313 [cond-mat.str-el]}
  \BibitemShut {NoStop}%
\bibitem [{\citenamefont {{Morimoto}}\ \emph {et~al.}(2015)\citenamefont
  {{Morimoto}}, \citenamefont {{Furusaki}},\ and\ \citenamefont
  {{Mudry}}}]{Morimoto1505.06341}%
  \BibitemOpen
  \bibfield  {author} {\bibinfo {author} {\bibfnamefont {T.}~\bibnamefont
  {{Morimoto}}}, \bibinfo {author} {\bibfnamefont {A.}~\bibnamefont
  {{Furusaki}}},\ and\ \bibinfo {author} {\bibfnamefont {C.}~\bibnamefont
  {{Mudry}}},\ }\bibfield  {title} {\bibinfo {title} {{Breakdown of the
  topological classification Z for gapped phases of noninteracting fermions by
  quartic interactions}},\ }\href {https://doi.org/10.1103/PhysRevB.92.125104}
  {\bibfield  {journal} {\bibinfo  {journal} {\prb}\ }\textbf {\bibinfo
  {volume} {92}},\ \bibinfo {eid} {125104} (\bibinfo {year} {2015})},\ \Eprint
  {https://arxiv.org/abs/1505.06341} {arXiv:1505.06341 [cond-mat.str-el]}
  \BibitemShut {NoStop}%
\bibitem [{\citenamefont {{Kapustin}}\ \emph {et~al.}(2015)\citenamefont
  {{Kapustin}}, \citenamefont {{Thorngren}}, \citenamefont {{Turzillo}},\ and\
  \citenamefont {{Wang}}}]{Kapustin2015Fermionic}%
  \BibitemOpen
  \bibfield  {author} {\bibinfo {author} {\bibfnamefont {A.}~\bibnamefont
  {{Kapustin}}}, \bibinfo {author} {\bibfnamefont {R.}~\bibnamefont
  {{Thorngren}}}, \bibinfo {author} {\bibfnamefont {A.}~\bibnamefont
  {{Turzillo}}},\ and\ \bibinfo {author} {\bibfnamefont {Z.}~\bibnamefont
  {{Wang}}},\ }\bibfield  {title} {\bibinfo {title} {{Fermionic symmetry
  protected topological phases and cobordisms}},\ }\href
  {https://doi.org/10.1007/JHEP12(2015)052} {\bibfield  {journal} {\bibinfo
  {journal} {Journal of High Energy Physics}\ }\textbf {\bibinfo {volume}
  {2015}},\ \bibinfo {eid} {52} (\bibinfo {year} {2015})},\ \Eprint
  {https://arxiv.org/abs/1406.7329} {arXiv:1406.7329 [cond-mat.str-el]}
  \BibitemShut {NoStop}%
\bibitem [{\citenamefont {{Freed}}\ and\ \citenamefont
  {{Hopkins}}(2016)}]{Freed2016Reflection}%
  \BibitemOpen
  \bibfield  {author} {\bibinfo {author} {\bibfnamefont {D.~S.}\ \bibnamefont
  {{Freed}}}\ and\ \bibinfo {author} {\bibfnamefont {M.~J.}\ \bibnamefont
  {{Hopkins}}},\ }\bibfield  {title} {\bibinfo {title} {{Reflection positivity
  and invertible topological phases}},\ }\href@noop {} {\bibfield  {journal}
  {\bibinfo  {journal} {arXiv e-prints}\ ,\ \bibinfo {eid} {arXiv:1604.06527}}
  (\bibinfo {year} {2016})},\ \Eprint {https://arxiv.org/abs/1604.06527}
  {arXiv:1604.06527 [hep-th]} \BibitemShut {NoStop}%
\bibitem [{\citenamefont {{Gaiotto}}\ and\ \citenamefont
  {{Kapustin}}(2016)}]{Gaiotto2016Spin}%
  \BibitemOpen
  \bibfield  {author} {\bibinfo {author} {\bibfnamefont {D.}~\bibnamefont
  {{Gaiotto}}}\ and\ \bibinfo {author} {\bibfnamefont {A.}~\bibnamefont
  {{Kapustin}}},\ }\bibfield  {title} {\bibinfo {title} {{Spin TQFTs and
  fermionic phases of matter}},\ }\href
  {https://doi.org/10.1142/S0217751X16450445} {\bibfield  {journal} {\bibinfo
  {journal} {International Journal of Modern Physics A}\ }\textbf {\bibinfo
  {volume} {31}},\ \bibinfo {eid} {1645044-184} (\bibinfo {year} {2016})},\
  \Eprint {https://arxiv.org/abs/1505.05856} {arXiv:1505.05856
  [cond-mat.str-el]} \BibitemShut {NoStop}%
\bibitem [{\citenamefont {{Wen}}(2017)}]{WenZoo1610.03911}%
  \BibitemOpen
  \bibfield  {author} {\bibinfo {author} {\bibfnamefont {X.-G.}\ \bibnamefont
  {{Wen}}},\ }\bibfield  {title} {\bibinfo {title} {Colloquium: Zoo of
  quantum-topological phases of matter},\ }\href
  {https://doi.org/10.1103/RevModPhys.89.041004} {\bibfield  {journal}
  {\bibinfo  {journal} {Rev. Mod. Phys.}\ }\textbf {\bibinfo {volume} {89}},\
  \bibinfo {pages} {041004} (\bibinfo {year} {2017})},\ \Eprint
  {https://arxiv.org/abs/1610.03911} {arXiv:1610.03911 [cond-mat.str-el]}
  \BibitemShut {NoStop}%
\bibitem [{\citenamefont {{Wang}}\ and\ \citenamefont
  {{Gu}}(2018{\natexlab{a}})}]{Wang1703.10937}%
  \BibitemOpen
  \bibfield  {author} {\bibinfo {author} {\bibfnamefont {Q.-R.}\ \bibnamefont
  {{Wang}}}\ and\ \bibinfo {author} {\bibfnamefont {Z.-C.}\ \bibnamefont
  {{Gu}}},\ }\bibfield  {title} {\bibinfo {title} {{Towards a complete
  classification of fermionic symmetry protected topological phases in 3D and a
  general group supercohomology theory}},\ }\href
  {https://doi.org/10.1103/PhysRevX.8.011055} {\bibfield  {journal} {\bibinfo
  {journal} {Phys. Rev. X}\ }\textbf {\bibinfo {volume} {8}},\ \bibinfo {eid}
  {arXiv:1703.10937} (\bibinfo {year} {2018}{\natexlab{a}})},\ \Eprint
  {https://arxiv.org/abs/1703.10937} {arXiv:1703.10937 [cond-mat.str-el]}
  \BibitemShut {NoStop}%
\bibitem [{\citenamefont {{Kapustin}}\ and\ \citenamefont
  {{Thorngren}}(2017)}]{Kapustin2017Fermionic}%
  \BibitemOpen
  \bibfield  {author} {\bibinfo {author} {\bibfnamefont {A.}~\bibnamefont
  {{Kapustin}}}\ and\ \bibinfo {author} {\bibfnamefont {R.}~\bibnamefont
  {{Thorngren}}},\ }\bibfield  {title} {\bibinfo {title} {{Fermionic SPT phases
  in higher dimensions and bosonization}},\ }\href
  {https://doi.org/10.1007/JHEP10(2017)080} {\bibfield  {journal} {\bibinfo
  {journal} {Journal of High Energy Physics}\ }\textbf {\bibinfo {volume}
  {2017}},\ \bibinfo {eid} {80} (\bibinfo {year} {2017})},\ \Eprint
  {https://arxiv.org/abs/1701.08264} {arXiv:1701.08264 [cond-mat.str-el]}
  \BibitemShut {NoStop}%
\bibitem [{\citenamefont {{Wang}}\ \emph {et~al.}(2018)\citenamefont {{Wang}},
  \citenamefont {{Ohmori}}, \citenamefont {{Putrov}}, \citenamefont {{Zheng}},
  \citenamefont {{Wan}}, \citenamefont {{Guo}}, \citenamefont {{Lin}},
  \citenamefont {{Gao}},\ and\ \citenamefont {{Yau}}}]{Wang2018Tunneling}%
  \BibitemOpen
  \bibfield  {author} {\bibinfo {author} {\bibfnamefont {J.}~\bibnamefont
  {{Wang}}}, \bibinfo {author} {\bibfnamefont {K.}~\bibnamefont {{Ohmori}}},
  \bibinfo {author} {\bibfnamefont {P.}~\bibnamefont {{Putrov}}}, \bibinfo
  {author} {\bibfnamefont {Y.}~\bibnamefont {{Zheng}}}, \bibinfo {author}
  {\bibfnamefont {Z.}~\bibnamefont {{Wan}}}, \bibinfo {author} {\bibfnamefont
  {M.}~\bibnamefont {{Guo}}}, \bibinfo {author} {\bibfnamefont
  {H.}~\bibnamefont {{Lin}}}, \bibinfo {author} {\bibfnamefont
  {P.}~\bibnamefont {{Gao}}},\ and\ \bibinfo {author} {\bibfnamefont {S.-T.}\
  \bibnamefont {{Yau}}},\ }\bibfield  {title} {\bibinfo {title} {{Tunneling
  topological vacua via extended operators: (Spin-)TQFT spectra and boundary
  deconfinement in various dimensions}},\ }\href
  {https://doi.org/10.1093/ptep/pty051} {\bibfield  {journal} {\bibinfo
  {journal} {Progress of Theoretical and Experimental Physics}\ }\textbf
  {\bibinfo {volume} {2018}},\ \bibinfo {eid} {053A01} (\bibinfo {year}
  {2018})},\ \Eprint {https://arxiv.org/abs/1801.05416} {arXiv:1801.05416
  [cond-mat.str-el]} \BibitemShut {NoStop}%
\bibitem [{\citenamefont {{Wang}}\ and\ \citenamefont
  {{Gu}}(2018{\natexlab{b}})}]{Wang2018Construction}%
  \BibitemOpen
  \bibfield  {author} {\bibinfo {author} {\bibfnamefont {Q.-R.}\ \bibnamefont
  {{Wang}}}\ and\ \bibinfo {author} {\bibfnamefont {Z.-C.}\ \bibnamefont
  {{Gu}}},\ }\bibfield  {title} {\bibinfo {title} {{Construction and
  classification of symmetry protected topological phases in interacting
  fermion systems}},\ }\href@noop {} {\bibfield  {journal} {\bibinfo  {journal}
  {arXiv e-prints}\ ,\ \bibinfo {eid} {arXiv:1811.00536}} (\bibinfo {year}
  {2018}{\natexlab{b}})},\ \Eprint {https://arxiv.org/abs/1811.00536}
  {arXiv:1811.00536 [cond-mat.str-el]} \BibitemShut {NoStop}%
\bibitem [{\citenamefont {{Gaiotto}}\ and\ \citenamefont
  {{Johnson-Freyd}}(2019)}]{Gaiotto2019Symmetry}%
  \BibitemOpen
  \bibfield  {author} {\bibinfo {author} {\bibfnamefont {D.}~\bibnamefont
  {{Gaiotto}}}\ and\ \bibinfo {author} {\bibfnamefont {T.}~\bibnamefont
  {{Johnson-Freyd}}},\ }\bibfield  {title} {\bibinfo {title} {{Symmetry
  protected topological phases and generalized cohomology}},\ }\href
  {https://doi.org/10.1007/JHEP05(2019)007} {\bibfield  {journal} {\bibinfo
  {journal} {Journal of High Energy Physics}\ }\textbf {\bibinfo {volume}
  {2019}},\ \bibinfo {eid} {7} (\bibinfo {year} {2019})},\ \Eprint
  {https://arxiv.org/abs/1712.07950} {arXiv:1712.07950 [hep-th]} \BibitemShut
  {NoStop}%
\bibitem [{\citenamefont {{Tong}}\ and\ \citenamefont
  {{Turner}}(2019)}]{Tong1906.07199}%
  \BibitemOpen
  \bibfield  {author} {\bibinfo {author} {\bibfnamefont {D.}~\bibnamefont
  {{Tong}}}\ and\ \bibinfo {author} {\bibfnamefont {C.}~\bibnamefont
  {{Turner}}},\ }\bibfield  {title} {\bibinfo {title} {{Notes on 8 Majorana
  Fermions}},\ }\href@noop {} {\bibfield  {journal} {\bibinfo  {journal} {arXiv
  e-prints}\ ,\ \bibinfo {eid} {arXiv:1906.07199}} (\bibinfo {year} {2019})},\
  \Eprint {https://arxiv.org/abs/1906.07199} {arXiv:1906.07199 [hep-th]}
  \BibitemShut {NoStop}%
\bibitem [{\citenamefont {{Lan}}\ \emph {et~al.}(2019)\citenamefont {{Lan}},
  \citenamefont {{Zhu}},\ and\ \citenamefont {{Wen}}}]{Lan2019Fermion}%
  \BibitemOpen
  \bibfield  {author} {\bibinfo {author} {\bibfnamefont {T.}~\bibnamefont
  {{Lan}}}, \bibinfo {author} {\bibfnamefont {C.}~\bibnamefont {{Zhu}}},\ and\
  \bibinfo {author} {\bibfnamefont {X.-G.}\ \bibnamefont {{Wen}}},\ }\bibfield
  {title} {\bibinfo {title} {{Fermion decoration construction of
  symmetry-protected trivial order for fermion systems with any symmetry and in
  any dimension}},\ }\href {https://doi.org/10.1103/PhysRevB.100.235141}
  {\bibfield  {journal} {\bibinfo  {journal} {\prb}\ }\textbf {\bibinfo
  {volume} {100}},\ \bibinfo {eid} {235141} (\bibinfo {year} {2019})},\ \Eprint
  {https://arxiv.org/abs/1809.01112} {arXiv:1809.01112 [cond-mat.str-el]}
  \BibitemShut {NoStop}%
\bibitem [{\citenamefont {{Guo}}\ \emph {et~al.}(2020)\citenamefont {{Guo}},
  \citenamefont {{Ohmori}}, \citenamefont {{Putrov}}, \citenamefont {{Wan}},\
  and\ \citenamefont {{Wang}}}]{Guo2020Fermionic}%
  \BibitemOpen
  \bibfield  {author} {\bibinfo {author} {\bibfnamefont {M.}~\bibnamefont
  {{Guo}}}, \bibinfo {author} {\bibfnamefont {K.}~\bibnamefont {{Ohmori}}},
  \bibinfo {author} {\bibfnamefont {P.}~\bibnamefont {{Putrov}}}, \bibinfo
  {author} {\bibfnamefont {Z.}~\bibnamefont {{Wan}}},\ and\ \bibinfo {author}
  {\bibfnamefont {J.}~\bibnamefont {{Wang}}},\ }\bibfield  {title} {\bibinfo
  {title} {{Fermionic Finite-Group Gauge Theories and Interacting
  Symmetric/Crystalline Orders via Cobordisms}},\ }\href
  {https://doi.org/10.1007/s00220-019-03671-6} {\bibfield  {journal} {\bibinfo
  {journal} {Communications in Mathematical Physics}\ }\textbf {\bibinfo
  {volume} {376}},\ \bibinfo {pages} {1073} (\bibinfo {year} {2020})},\ \Eprint
  {https://arxiv.org/abs/1812.11959} {arXiv:1812.11959 [hep-th]} \BibitemShut
  {NoStop}%
\bibitem [{\citenamefont {{Ouyang}}\ \emph {et~al.}(2020)\citenamefont
  {{Ouyang}}, \citenamefont {{Wang}}, \citenamefont {{Gu}},\ and\ \citenamefont
  {{Qi}}}]{Ouyang2020Computing}%
  \BibitemOpen
  \bibfield  {author} {\bibinfo {author} {\bibfnamefont {Y.}~\bibnamefont
  {{Ouyang}}}, \bibinfo {author} {\bibfnamefont {Q.-R.}\ \bibnamefont
  {{Wang}}}, \bibinfo {author} {\bibfnamefont {Z.-C.}\ \bibnamefont {{Gu}}},\
  and\ \bibinfo {author} {\bibfnamefont {Y.}~\bibnamefont {{Qi}}},\ }\bibfield
  {title} {\bibinfo {title} {{Computing classification of interacting fermionic
  symmetry-protected topological phases using topological invariants}},\
  }\href@noop {} {\bibfield  {journal} {\bibinfo  {journal} {arXiv e-prints}\
  ,\ \bibinfo {eid} {arXiv:2005.06572}} (\bibinfo {year} {2020})},\ \Eprint
  {https://arxiv.org/abs/2005.06572} {arXiv:2005.06572 [cond-mat.str-el]}
  \BibitemShut {NoStop}%
\bibitem [{\citenamefont {{Aasen}}\ \emph {et~al.}(2021)\citenamefont
  {{Aasen}}, \citenamefont {{Bonderson}},\ and\ \citenamefont
  {{Knapp}}}]{Aasen2109.10911}%
  \BibitemOpen
  \bibfield  {author} {\bibinfo {author} {\bibfnamefont {D.}~\bibnamefont
  {{Aasen}}}, \bibinfo {author} {\bibfnamefont {P.}~\bibnamefont
  {{Bonderson}}},\ and\ \bibinfo {author} {\bibfnamefont {C.}~\bibnamefont
  {{Knapp}}},\ }\bibfield  {title} {\bibinfo {title} {{Characterization and
  Classification of Fermionic Symmetry Enriched Topological Phases}},\ }\href
  {https://doi.org/10.48550/arXiv.2109.10911} {\bibfield  {journal} {\bibinfo
  {journal} {arXiv e-prints}\ ,\ \bibinfo {eid} {arXiv:2109.10911}} (\bibinfo
  {year} {2021})},\ \Eprint {https://arxiv.org/abs/2109.10911}
  {arXiv:2109.10911 [cond-mat.str-el]} \BibitemShut {NoStop}%
\bibitem [{\citenamefont {{Barkeshli}}\ \emph {et~al.}(2022)\citenamefont
  {{Barkeshli}}, \citenamefont {{Chen}}, \citenamefont {{Hsin}},\ and\
  \citenamefont {{Manjunath}}}]{Barkeshli2109.11039}%
  \BibitemOpen
  \bibfield  {author} {\bibinfo {author} {\bibfnamefont {M.}~\bibnamefont
  {{Barkeshli}}}, \bibinfo {author} {\bibfnamefont {Y.-A.}\ \bibnamefont
  {{Chen}}}, \bibinfo {author} {\bibfnamefont {P.-S.}\ \bibnamefont {{Hsin}}},\
  and\ \bibinfo {author} {\bibfnamefont {N.}~\bibnamefont {{Manjunath}}},\
  }\bibfield  {title} {\bibinfo {title} {{Classification of (2 +1 )D invertible
  fermionic topological phases with symmetry}},\ }\href
  {https://doi.org/10.1103/PhysRevB.105.235143} {\bibfield  {journal} {\bibinfo
   {journal} {\prb}\ }\textbf {\bibinfo {volume} {105}},\ \bibinfo {eid}
  {235143} (\bibinfo {year} {2022})},\ \Eprint
  {https://arxiv.org/abs/2109.11039} {arXiv:2109.11039 [cond-mat.str-el]}
  \BibitemShut {NoStop}%
\bibitem [{\citenamefont {{Hasenfratz}}(2022)}]{Hasenfratz2204.04801}%
  \BibitemOpen
  \bibfield  {author} {\bibinfo {author} {\bibfnamefont {A.}~\bibnamefont
  {{Hasenfratz}}},\ }\bibfield  {title} {\bibinfo {title} {{Emergent strongly
  coupled ultraviolet fixed point in four dimensions with 8 K{\"a}hler-Dirac
  fermions}},\ }\href@noop {} {\bibfield  {journal} {\bibinfo  {journal} {arXiv
  e-prints}\ ,\ \bibinfo {eid} {arXiv:2204.04801}} (\bibinfo {year} {2022})},\
  \Eprint {https://arxiv.org/abs/2204.04801} {arXiv:2204.04801 [hep-lat]}
  \BibitemShut {NoStop}%
\bibitem [{\citenamefont {{Manjunath}}\ \emph {et~al.}(2022)\citenamefont
  {{Manjunath}}, \citenamefont {{Calvera}},\ and\ \citenamefont
  {{Barkeshli}}}]{Manjunath2210.02452}%
  \BibitemOpen
  \bibfield  {author} {\bibinfo {author} {\bibfnamefont {N.}~\bibnamefont
  {{Manjunath}}}, \bibinfo {author} {\bibfnamefont {V.}~\bibnamefont
  {{Calvera}}},\ and\ \bibinfo {author} {\bibfnamefont {M.}~\bibnamefont
  {{Barkeshli}}},\ }\bibfield  {title} {\bibinfo {title} {{Non-perturbative
  constraints from symmetry and chirality on Majorana zero modes and defect
  quantum numbers in (2+1)D}},\ }\href
  {https://doi.org/10.48550/arXiv.2210.02452} {\bibfield  {journal} {\bibinfo
  {journal} {arXiv e-prints}\ ,\ \bibinfo {eid} {arXiv:2210.02452}} (\bibinfo
  {year} {2022})},\ \Eprint {https://arxiv.org/abs/2210.02452}
  {arXiv:2210.02452 [cond-mat.str-el]} \BibitemShut {NoStop}%
\bibitem [{\citenamefont {{Zhang}}\ \emph
  {et~al.}(2022{\natexlab{a}})\citenamefont {{Zhang}}, \citenamefont
  {{Manjunath}}, \citenamefont {{Nambiar}},\ and\ \citenamefont
  {{Barkeshli}}}]{Zhang2211.09127}%
  \BibitemOpen
  \bibfield  {author} {\bibinfo {author} {\bibfnamefont {Y.}~\bibnamefont
  {{Zhang}}}, \bibinfo {author} {\bibfnamefont {N.}~\bibnamefont
  {{Manjunath}}}, \bibinfo {author} {\bibfnamefont {G.}~\bibnamefont
  {{Nambiar}}},\ and\ \bibinfo {author} {\bibfnamefont {M.}~\bibnamefont
  {{Barkeshli}}},\ }\bibfield  {title} {\bibinfo {title} {{Quantized charge
  polarization as a many-body invariant in (2+1)D crystalline topological
  states and Hofstadter butterflies}},\ }\href
  {https://doi.org/10.48550/arXiv.2211.09127} {\bibfield  {journal} {\bibinfo
  {journal} {arXiv e-prints}\ ,\ \bibinfo {eid} {arXiv:2211.09127}} (\bibinfo
  {year} {2022}{\natexlab{a}})},\ \Eprint {https://arxiv.org/abs/2211.09127}
  {arXiv:2211.09127 [cond-mat.str-el]} \BibitemShut {NoStop}%
\bibitem [{\citenamefont {{Zhang}}\ \emph
  {et~al.}(2022{\natexlab{b}})\citenamefont {{Zhang}}, \citenamefont
  {{Manjunath}}, \citenamefont {{Nambiar}},\ and\ \citenamefont
  {{Barkeshli}}}]{Zhang2204.05320}%
  \BibitemOpen
  \bibfield  {author} {\bibinfo {author} {\bibfnamefont {Y.}~\bibnamefont
  {{Zhang}}}, \bibinfo {author} {\bibfnamefont {N.}~\bibnamefont
  {{Manjunath}}}, \bibinfo {author} {\bibfnamefont {G.}~\bibnamefont
  {{Nambiar}}},\ and\ \bibinfo {author} {\bibfnamefont {M.}~\bibnamefont
  {{Barkeshli}}},\ }\bibfield  {title} {\bibinfo {title} {{Fractional
  Disclination Charge and Discrete Shift in the Hofstadter Butterfly}},\ }\href
  {https://doi.org/10.1103/PhysRevLett.129.275301} {\bibfield  {journal}
  {\bibinfo  {journal} {\prl}\ }\textbf {\bibinfo {volume} {129}},\ \bibinfo
  {eid} {275301} (\bibinfo {year} {2022}{\natexlab{b}})},\ \Eprint
  {https://arxiv.org/abs/2204.05320} {arXiv:2204.05320 [cond-mat.str-el]}
  \BibitemShut {NoStop}%
\bibitem [{\citenamefont {{Ryu}}\ and\ \citenamefont
  {{Zhang}}(2012)}]{Ryu:2012ph}%
  \BibitemOpen
  \bibfield  {author} {\bibinfo {author} {\bibfnamefont {S.}~\bibnamefont
  {{Ryu}}}\ and\ \bibinfo {author} {\bibfnamefont {S.-C.}\ \bibnamefont
  {{Zhang}}},\ }\bibfield  {title} {\bibinfo {title} {{Interacting topological
  phases and modular invariance}},\ }\href
  {https://doi.org/10.1103/PhysRevB.85.245132} {\bibfield  {journal} {\bibinfo
  {journal} {\prb}\ }\textbf {\bibinfo {volume} {85}},\ \bibinfo {eid} {245132}
  (\bibinfo {year} {2012})},\ \Eprint {https://arxiv.org/abs/1202.4484}
  {arXiv:1202.4484 [cond-mat.str-el]} \BibitemShut {NoStop}%
\bibitem [{\citenamefont {{Qi}}(2013)}]{Qi:2013qe}%
  \BibitemOpen
  \bibfield  {author} {\bibinfo {author} {\bibfnamefont {X.-L.}\ \bibnamefont
  {{Qi}}},\ }\bibfield  {title} {\bibinfo {title} {{A new class of (2 +
  1)-dimensional topological superconductors with $\mathbb{Z}_8$ topological
  classification}},\ }\href {https://doi.org/10.1088/1367-2630/15/6/065002}
  {\bibfield  {journal} {\bibinfo  {journal} {New Journal of Physics}\ }\textbf
  {\bibinfo {volume} {15}},\ \bibinfo {eid} {065002} (\bibinfo {year}
  {2013})},\ \Eprint {https://arxiv.org/abs/1202.3983} {arXiv:1202.3983
  [cond-mat.str-el]} \BibitemShut {NoStop}%
\bibitem [{\citenamefont {{Yao}}\ and\ \citenamefont
  {{Ryu}}(2013)}]{Yao:2013yg}%
  \BibitemOpen
  \bibfield  {author} {\bibinfo {author} {\bibfnamefont {H.}~\bibnamefont
  {{Yao}}}\ and\ \bibinfo {author} {\bibfnamefont {S.}~\bibnamefont {{Ryu}}},\
  }\bibfield  {title} {\bibinfo {title} {{Interaction effect on topological
  classification of superconductors in two dimensions}},\ }\href
  {https://doi.org/10.1103/PhysRevB.88.064507} {\bibfield  {journal} {\bibinfo
  {journal} {\prb}\ }\textbf {\bibinfo {volume} {88}},\ \bibinfo {eid} {064507}
  (\bibinfo {year} {2013})},\ \Eprint {https://arxiv.org/abs/1202.5805}
  {arXiv:1202.5805 [cond-mat.str-el]} \BibitemShut {NoStop}%
\bibitem [{\citenamefont {{Wang}}\ and\ \citenamefont
  {{Senthil}}(2014)}]{Wang2014Interacting}%
  \BibitemOpen
  \bibfield  {author} {\bibinfo {author} {\bibfnamefont {C.}~\bibnamefont
  {{Wang}}}\ and\ \bibinfo {author} {\bibfnamefont {T.}~\bibnamefont
  {{Senthil}}},\ }\bibfield  {title} {\bibinfo {title} {{Interacting fermionic
  topological insulators/superconductors in three dimensions}},\ }\href
  {https://doi.org/10.1103/PhysRevB.89.195124} {\bibfield  {journal} {\bibinfo
  {journal} {\prb}\ }\textbf {\bibinfo {volume} {89}},\ \bibinfo {eid} {195124}
  (\bibinfo {year} {2014})},\ \Eprint {https://arxiv.org/abs/1401.1142}
  {arXiv:1401.1142 [cond-mat.str-el]} \BibitemShut {NoStop}%
\bibitem [{\citenamefont {{Gu}}\ and\ \citenamefont
  {{Levin}}(2014)}]{Gu:2014tw}%
  \BibitemOpen
  \bibfield  {author} {\bibinfo {author} {\bibfnamefont {Z.-C.}\ \bibnamefont
  {{Gu}}}\ and\ \bibinfo {author} {\bibfnamefont {M.}~\bibnamefont {{Levin}}},\
  }\bibfield  {title} {\bibinfo {title} {{Effect of interactions on
  two-dimensional fermionic symmetry-protected topological phases with Z$_{2}$
  symmetry}},\ }\href {https://doi.org/10.1103/PhysRevB.89.201113} {\bibfield
  {journal} {\bibinfo  {journal} {\prb}\ }\textbf {\bibinfo {volume} {89}},\
  \bibinfo {eid} {201113} (\bibinfo {year} {2014})},\ \Eprint
  {https://arxiv.org/abs/1304.4569} {arXiv:1304.4569 [cond-mat.str-el]}
  \BibitemShut {NoStop}%
\bibitem [{\citenamefont {{Metlitski}}\ \emph {et~al.}(2014)\citenamefont
  {{Metlitski}}, \citenamefont {{Fidkowski}}, \citenamefont {{Chen}},\ and\
  \citenamefont {{Vishwanath}}}]{Metlitski2014Interaction}%
  \BibitemOpen
  \bibfield  {author} {\bibinfo {author} {\bibfnamefont {M.~A.}\ \bibnamefont
  {{Metlitski}}}, \bibinfo {author} {\bibfnamefont {L.}~\bibnamefont
  {{Fidkowski}}}, \bibinfo {author} {\bibfnamefont {X.}~\bibnamefont
  {{Chen}}},\ and\ \bibinfo {author} {\bibfnamefont {A.}~\bibnamefont
  {{Vishwanath}}},\ }\bibfield  {title} {\bibinfo {title} {{Interaction effects
  on 3D topological superconductors: surface topological order from vortex
  condensation, the 16 fold way and fermionic Kramers doublets}},\ }\href@noop
  {} {\bibfield  {journal} {\bibinfo  {journal} {arXiv e-prints}\ ,\ \bibinfo
  {eid} {arXiv:1406.3032}} (\bibinfo {year} {2014})},\ \Eprint
  {https://arxiv.org/abs/1406.3032} {arXiv:1406.3032 [cond-mat.str-el]}
  \BibitemShut {NoStop}%
\bibitem [{\citenamefont {{You}}\ and\ \citenamefont
  {{Xu}}(2014)}]{You:2014ho}%
  \BibitemOpen
  \bibfield  {author} {\bibinfo {author} {\bibfnamefont {Y.-Z.}\ \bibnamefont
  {{You}}}\ and\ \bibinfo {author} {\bibfnamefont {C.}~\bibnamefont {{Xu}}},\
  }\bibfield  {title} {\bibinfo {title} {{Symmetry-protected topological states
  of interacting fermions and bosons}},\ }\href
  {https://doi.org/10.1103/PhysRevB.90.245120} {\bibfield  {journal} {\bibinfo
  {journal} {\prb}\ }\textbf {\bibinfo {volume} {90}},\ \bibinfo {eid} {245120}
  (\bibinfo {year} {2014})},\ \Eprint {https://arxiv.org/abs/1409.0168}
  {arXiv:1409.0168 [cond-mat.str-el]} \BibitemShut {NoStop}%
\bibitem [{\citenamefont {{Yoshida}}\ and\ \citenamefont
  {{Furusaki}}(2015)}]{Yoshida:2015aj}%
  \BibitemOpen
  \bibfield  {author} {\bibinfo {author} {\bibfnamefont {T.}~\bibnamefont
  {{Yoshida}}}\ and\ \bibinfo {author} {\bibfnamefont {A.}~\bibnamefont
  {{Furusaki}}},\ }\bibfield  {title} {\bibinfo {title} {{Correlation effects
  on topological crystalline insulators}},\ }\href
  {https://doi.org/10.1103/PhysRevB.92.085114} {\bibfield  {journal} {\bibinfo
  {journal} {\prb}\ }\textbf {\bibinfo {volume} {92}},\ \bibinfo {eid} {085114}
  (\bibinfo {year} {2015})},\ \Eprint {https://arxiv.org/abs/1505.06598}
  {arXiv:1505.06598 [cond-mat.str-el]} \BibitemShut {NoStop}%
\bibitem [{\citenamefont {{Gu}}\ and\ \citenamefont {{Qi}}(2015)}]{Gu:2015cy}%
  \BibitemOpen
  \bibfield  {author} {\bibinfo {author} {\bibfnamefont {Y.}~\bibnamefont
  {{Gu}}}\ and\ \bibinfo {author} {\bibfnamefont {X.-L.}\ \bibnamefont
  {{Qi}}},\ }\bibfield  {title} {\bibinfo {title} {{Axion field theory approach
  and the classification of interacting topological superconductors}},\
  }\href@noop {} {\bibfield  {journal} {\bibinfo  {journal} {ArXiv e-prints}\ }
  (\bibinfo {year} {2015})},\ \Eprint {https://arxiv.org/abs/1512.04919}
  {arXiv:1512.04919 [cond-mat.supr-con]} \BibitemShut {NoStop}%
\bibitem [{\citenamefont {{Song}}\ and\ \citenamefont
  {{Schnyder}}(2017)}]{Song1609.07469}%
  \BibitemOpen
  \bibfield  {author} {\bibinfo {author} {\bibfnamefont {X.-Y.}\ \bibnamefont
  {{Song}}}\ and\ \bibinfo {author} {\bibfnamefont {A.~P.}\ \bibnamefont
  {{Schnyder}}},\ }\bibfield  {title} {\bibinfo {title} {{Interaction effects
  on the classification of crystalline topological insulators and
  superconductors}},\ }\href {https://doi.org/10.1103/PhysRevB.95.195108}
  {\bibfield  {journal} {\bibinfo  {journal} {\prb}\ }\textbf {\bibinfo
  {volume} {95}},\ \bibinfo {eid} {195108} (\bibinfo {year} {2017})},\ \Eprint
  {https://arxiv.org/abs/1609.07469} {arXiv:1609.07469 [cond-mat.str-el]}
  \BibitemShut {NoStop}%
\bibitem [{\citenamefont {{Queiroz}}\ \emph {et~al.}(2016)\citenamefont
  {{Queiroz}}, \citenamefont {{Khalaf}},\ and\ \citenamefont
  {{Stern}}}]{Queiroz:2016se}%
  \BibitemOpen
  \bibfield  {author} {\bibinfo {author} {\bibfnamefont {R.}~\bibnamefont
  {{Queiroz}}}, \bibinfo {author} {\bibfnamefont {E.}~\bibnamefont
  {{Khalaf}}},\ and\ \bibinfo {author} {\bibfnamefont {A.}~\bibnamefont
  {{Stern}}},\ }\bibfield  {title} {\bibinfo {title} {{Dimensional Hierarchy of
  Fermionic Interacting Topological Phases}},\ }\href
  {https://doi.org/10.1103/PhysRevLett.117.206405} {\bibfield  {journal}
  {\bibinfo  {journal} {Physical Review Letters}\ }\textbf {\bibinfo {volume}
  {117}},\ \bibinfo {eid} {206405} (\bibinfo {year} {2016})},\ \Eprint
  {https://arxiv.org/abs/1601.01596} {arXiv:1601.01596 [cond-mat.str-el]}
  \BibitemShut {NoStop}%
\bibitem [{\citenamefont {{Wen}}(2013)}]{Wen:2013kr}%
  \BibitemOpen
  \bibfield  {author} {\bibinfo {author} {\bibfnamefont {X.-G.}\ \bibnamefont
  {{Wen}}},\ }\bibfield  {title} {\bibinfo {title} {{A Lattice Non-Perturbative
  Definition of an SO(10) Chiral Gauge Theory and Its Induced Standard
  Model}},\ }\href {https://doi.org/10.1088/0256-307X/30/11/111101} {\bibfield
  {journal} {\bibinfo  {journal} {Chinese Physics Letters}\ }\textbf {\bibinfo
  {volume} {30}},\ \bibinfo {eid} {111101} (\bibinfo {year} {2013})},\ \Eprint
  {https://arxiv.org/abs/1305.1045} {arXiv:1305.1045 [hep-lat]} \BibitemShut
  {NoStop}%
\bibitem [{\citenamefont {{You}}\ \emph {et~al.}(2014)\citenamefont {{You}},
  \citenamefont {{BenTov}},\ and\ \citenamefont {{Xu}}}]{You:2014ow}%
  \BibitemOpen
  \bibfield  {author} {\bibinfo {author} {\bibfnamefont {Y.-Z.}\ \bibnamefont
  {{You}}}, \bibinfo {author} {\bibfnamefont {Y.}~\bibnamefont {{BenTov}}},\
  and\ \bibinfo {author} {\bibfnamefont {C.}~\bibnamefont {{Xu}}},\ }\bibfield
  {title} {\bibinfo {title} {{Interacting Topological Superconductors and
  possible Origin of $16n$ Chiral Fermions in the Standard Model}},\
  }\href@noop {} {\bibfield  {journal} {\bibinfo  {journal} {ArXiv e-prints}\ }
  (\bibinfo {year} {2014})},\ \Eprint {https://arxiv.org/abs/1402.4151}
  {arXiv:1402.4151 [cond-mat.str-el]} \BibitemShut {NoStop}%
\bibitem [{\citenamefont {{You}}\ and\ \citenamefont
  {{Xu}}(2015)}]{You:2015lj}%
  \BibitemOpen
  \bibfield  {author} {\bibinfo {author} {\bibfnamefont {Y.-Z.}\ \bibnamefont
  {{You}}}\ and\ \bibinfo {author} {\bibfnamefont {C.}~\bibnamefont {{Xu}}},\
  }\bibfield  {title} {\bibinfo {title} {{Interacting topological insulator and
  emergent grand unified theory}},\ }\href
  {https://doi.org/10.1103/PhysRevB.91.125147} {\bibfield  {journal} {\bibinfo
  {journal} {\prb}\ }\textbf {\bibinfo {volume} {91}},\ \bibinfo {eid} {125147}
  (\bibinfo {year} {2015})},\ \Eprint {https://arxiv.org/abs/1412.4784}
  {arXiv:1412.4784 [cond-mat.str-el]} \BibitemShut {NoStop}%
\bibitem [{\citenamefont {{BenTov}}(2015)}]{BenTov:2015lh}%
  \BibitemOpen
  \bibfield  {author} {\bibinfo {author} {\bibfnamefont {Y.}~\bibnamefont
  {{BenTov}}},\ }\bibfield  {title} {\bibinfo {title} {{Fermion masses without
  symmetry breaking in two spacetime dimensions}},\ }\href
  {https://doi.org/10.1007/JHEP07(2015)034} {\bibfield  {journal} {\bibinfo
  {journal} {Journal of High Energy Physics}\ }\textbf {\bibinfo {volume}
  {7}},\ \bibinfo {eid} {34} (\bibinfo {year} {2015})},\ \Eprint
  {https://arxiv.org/abs/1412.0154} {arXiv:1412.0154 [cond-mat.str-el]}
  \BibitemShut {NoStop}%
\bibitem [{\citenamefont {{DeMarco}}\ and\ \citenamefont
  {{Wen}}(2017)}]{DeMarco2017A-Novel}%
  \BibitemOpen
  \bibfield  {author} {\bibinfo {author} {\bibfnamefont {M.}~\bibnamefont
  {{DeMarco}}}\ and\ \bibinfo {author} {\bibfnamefont {X.-G.}\ \bibnamefont
  {{Wen}}},\ }\bibfield  {title} {\bibinfo {title} {{A Novel Non-Perturbative
  Lattice Regularization of an Anomaly-Free $1 + 1d$ Chiral $SU(2)$ Gauge
  Theory}},\ }\href@noop {} {\bibfield  {journal} {\bibinfo  {journal} {arXiv
  e-prints}\ ,\ \bibinfo {eid} {arXiv:1706.04648}} (\bibinfo {year} {2017})},\
  \Eprint {https://arxiv.org/abs/1706.04648} {arXiv:1706.04648 [hep-lat]}
  \BibitemShut {NoStop}%
\bibitem [{\citenamefont {{Wang}}\ and\ \citenamefont
  {{Wen}}(2018)}]{Wang2018A-Non-Perturbative}%
  \BibitemOpen
  \bibfield  {author} {\bibinfo {author} {\bibfnamefont {J.}~\bibnamefont
  {{Wang}}}\ and\ \bibinfo {author} {\bibfnamefont {X.-G.}\ \bibnamefont
  {{Wen}}},\ }\bibfield  {title} {\bibinfo {title} {{A Non-Perturbative
  Definition of the Standard Models}},\ }\href@noop {} {\bibfield  {journal}
  {\bibinfo  {journal} {arXiv e-prints}\ ,\ \bibinfo {eid} {arXiv:1809.11171}}
  (\bibinfo {year} {2018})},\ \Eprint {https://arxiv.org/abs/1809.11171}
  {arXiv:1809.11171 [hep-th]} \BibitemShut {NoStop}%
\bibitem [{\citenamefont {{Wang}}\ and\ \citenamefont
  {{Wen}}(2019)}]{Wang2019Solution}%
  \BibitemOpen
  \bibfield  {author} {\bibinfo {author} {\bibfnamefont {J.}~\bibnamefont
  {{Wang}}}\ and\ \bibinfo {author} {\bibfnamefont {X.-G.}\ \bibnamefont
  {{Wen}}},\ }\bibfield  {title} {\bibinfo {title} {{Solution to the 1 +1
  dimensional gauged chiral Fermion problem}},\ }\href
  {https://doi.org/10.1103/PhysRevD.99.111501} {\bibfield  {journal} {\bibinfo
  {journal} {\prd}\ }\textbf {\bibinfo {volume} {99}},\ \bibinfo {eid} {111501}
  (\bibinfo {year} {2019})},\ \Eprint {https://arxiv.org/abs/1807.05998}
  {arXiv:1807.05998 [hep-lat]} \BibitemShut {NoStop}%
\bibitem [{\citenamefont {{Kikukawa}}(2019)}]{Kikukawa2019Why-is-the-mission}%
  \BibitemOpen
  \bibfield  {author} {\bibinfo {author} {\bibfnamefont {Y.}~\bibnamefont
  {{Kikukawa}}},\ }\bibfield  {title} {\bibinfo {title} {{Why is the mission
  impossible? Decoupling the mirror Ginsparg-Wilson fermions in the lattice
  models for two-dimensional Abelian chiral gauge theories}},\ }\href
  {https://doi.org/10.1093/ptep/ptz055} {\bibfield  {journal} {\bibinfo
  {journal} {Progress of Theoretical and Experimental Physics}\ }\textbf
  {\bibinfo {volume} {2019}},\ \bibinfo {eid} {073B02} (\bibinfo {year}
  {2019})},\ \Eprint {https://arxiv.org/abs/1710.11101} {arXiv:1710.11101
  [hep-lat]} \BibitemShut {NoStop}%
\bibitem [{\citenamefont {{Razamat}}\ and\ \citenamefont
  {{Tong}}(2021)}]{Razamat2021Gapped}%
  \BibitemOpen
  \bibfield  {author} {\bibinfo {author} {\bibfnamefont {S.~S.}\ \bibnamefont
  {{Razamat}}}\ and\ \bibinfo {author} {\bibfnamefont {D.}~\bibnamefont
  {{Tong}}},\ }\bibfield  {title} {\bibinfo {title} {{Gapped Chiral
  Fermions}},\ }\href {https://doi.org/10.1103/PhysRevX.11.011063} {\bibfield
  {journal} {\bibinfo  {journal} {Physical Review X}\ }\textbf {\bibinfo
  {volume} {11}},\ \bibinfo {eid} {011063} (\bibinfo {year} {2021})},\ \Eprint
  {https://arxiv.org/abs/2009.05037} {arXiv:2009.05037 [hep-th]} \BibitemShut
  {NoStop}%
\bibitem [{\citenamefont {{Butt}}\ \emph
  {et~al.}(2021{\natexlab{b}})\citenamefont {{Butt}}, \citenamefont
  {{Catterall}}, \citenamefont {{Pradhan}},\ and\ \citenamefont
  {{Toga}}}]{Butt2021Anomalies}%
  \BibitemOpen
  \bibfield  {author} {\bibinfo {author} {\bibfnamefont {N.}~\bibnamefont
  {{Butt}}}, \bibinfo {author} {\bibfnamefont {S.}~\bibnamefont {{Catterall}}},
  \bibinfo {author} {\bibfnamefont {A.}~\bibnamefont {{Pradhan}}},\ and\
  \bibinfo {author} {\bibfnamefont {G.~C.}\ \bibnamefont {{Toga}}},\ }\bibfield
   {title} {\bibinfo {title} {{Anomalies and symmetric mass generation for
  K{\"a}hler-Dirac fermions}},\ }\href
  {https://doi.org/10.1103/PhysRevD.104.094504} {\bibfield  {journal} {\bibinfo
   {journal} {\prd}\ }\textbf {\bibinfo {volume} {104}},\ \bibinfo {eid}
  {094504} (\bibinfo {year} {2021}{\natexlab{b}})},\ \Eprint
  {https://arxiv.org/abs/2101.01026} {arXiv:2101.01026 [hep-th]} \BibitemShut
  {NoStop}%
\bibitem [{\citenamefont {Zeng}\ \emph {et~al.}(2022)\citenamefont {Zeng},
  \citenamefont {Zhu}, \citenamefont {Wang},\ and\ \citenamefont
  {You}}]{Zeng2022Symmetric}%
  \BibitemOpen
  \bibfield  {author} {\bibinfo {author} {\bibfnamefont {M.}~\bibnamefont
  {Zeng}}, \bibinfo {author} {\bibfnamefont {Z.}~\bibnamefont {Zhu}}, \bibinfo
  {author} {\bibfnamefont {J.}~\bibnamefont {Wang}},\ and\ \bibinfo {author}
  {\bibfnamefont {Y.-Z.}\ \bibnamefont {You}},\ }\bibfield  {title} {\bibinfo
  {title} {{Symmetric Mass Generation in the 1+1 Dimensional Chiral Fermion
  3-4-5-0 Model}},\ }\href {https://doi.org/10.1103/PhysRevLett.128.185301}
  {\bibfield  {journal} {\bibinfo  {journal} {Phys. Rev. Lett.}\ }\textbf
  {\bibinfo {volume} {128}},\ \bibinfo {pages} {185301} (\bibinfo {year}
  {2022})},\ \Eprint {https://arxiv.org/abs/2202.12355} {arXiv:2202.12355
  [cond-mat.str-el]} \BibitemShut {NoStop}%
\bibitem [{\citenamefont {Nielsen}\ and\ \citenamefont
  {Ninomiya}(1981)}]{Nielsen1981Absence}%
  \BibitemOpen
  \bibfield  {author} {\bibinfo {author} {\bibfnamefont {H.~B.}\ \bibnamefont
  {Nielsen}}\ and\ \bibinfo {author} {\bibfnamefont {M.}~\bibnamefont
  {Ninomiya}},\ }\bibfield  {title} {\bibinfo {title} {Absence of neutrinos on
  a lattice: (ii). intuitive topological proof},\ }\href
  {https://doi.org/https://doi.org/10.1016/0550-3213(81)90524-1} {\bibfield
  {journal} {\bibinfo  {journal} {Nuclear Physics B}\ }\textbf {\bibinfo
  {volume} {193}},\ \bibinfo {pages} {173} (\bibinfo {year}
  {1981})}\BibitemShut {NoStop}%
\bibitem [{\citenamefont {{Nielsen}}\ and\ \citenamefont
  {{Ninomiya}}(1981{\natexlab{a}})}]{Nielsen:1981we}%
  \BibitemOpen
  \bibfield  {author} {\bibinfo {author} {\bibfnamefont {H.~B.}\ \bibnamefont
  {{Nielsen}}}\ and\ \bibinfo {author} {\bibfnamefont {M.}~\bibnamefont
  {{Ninomiya}}},\ }\bibfield  {title} {\bibinfo {title} {{Absence of neutrinos
  on a lattice (I). Proof by homotopy theory}},\ }\href
  {https://doi.org/10.1016/0550-3213(81)90361-8} {\bibfield  {journal}
  {\bibinfo  {journal} {Nuclear Physics B}\ }\textbf {\bibinfo {volume}
  {185}},\ \bibinfo {pages} {20} (\bibinfo {year}
  {1981}{\natexlab{a}})}\BibitemShut {NoStop}%
\bibitem [{\citenamefont {{Nielsen}}\ and\ \citenamefont
  {{Ninomiya}}(1981{\natexlab{b}})}]{Nielsen:1981bc}%
  \BibitemOpen
  \bibfield  {author} {\bibinfo {author} {\bibfnamefont {H.~B.}\ \bibnamefont
  {{Nielsen}}}\ and\ \bibinfo {author} {\bibfnamefont {M.}~\bibnamefont
  {{Ninomiya}}},\ }\bibfield  {title} {\bibinfo {title} {{A no-go theorem for
  regularizing chiral fermions}},\ }\href
  {https://doi.org/10.1016/0370-2693(81)91026-1} {\bibfield  {journal}
  {\bibinfo  {journal} {Physics Letters B}\ }\textbf {\bibinfo {volume}
  {105}},\ \bibinfo {pages} {219} (\bibinfo {year}
  {1981}{\natexlab{b}})}\BibitemShut {NoStop}%
\bibitem [{\citenamefont {{Catterall}}\ and\ \citenamefont
  {{Pradhan}}(2022)}]{Catterall2201.00750}%
  \BibitemOpen
  \bibfield  {author} {\bibinfo {author} {\bibfnamefont {S.}~\bibnamefont
  {{Catterall}}}\ and\ \bibinfo {author} {\bibfnamefont {A.}~\bibnamefont
  {{Pradhan}}},\ }\bibfield  {title} {\bibinfo {title} {{Induced topological
  gravity and anomaly inflow from K{\"a}hler-Dirac fermions in odd
  dimensions}},\ }\href {https://doi.org/10.1103/PhysRevD.106.014509}
  {\bibfield  {journal} {\bibinfo  {journal} {\prd}\ }\textbf {\bibinfo
  {volume} {106}},\ \bibinfo {eid} {014509} (\bibinfo {year} {2022})},\ \Eprint
  {https://arxiv.org/abs/2201.00750} {arXiv:2201.00750 [hep-th]} \BibitemShut
  {NoStop}%
\bibitem [{\citenamefont {{Catterall}}(2023)}]{Catterall2209.03828}%
  \BibitemOpen
  \bibfield  {author} {\bibinfo {author} {\bibfnamefont {S.}~\bibnamefont
  {{Catterall}}},\ }\bibfield  {title} {\bibinfo {title} {{'t Hooft anomalies
  for staggered fermions}},\ }\href
  {https://doi.org/10.1103/PhysRevD.107.014501} {\bibfield  {journal} {\bibinfo
   {journal} {\prd}\ }\textbf {\bibinfo {volume} {107}},\ \bibinfo {eid}
  {014501} (\bibinfo {year} {2023})},\ \Eprint
  {https://arxiv.org/abs/2209.03828} {arXiv:2209.03828 [hep-lat]} \BibitemShut
  {NoStop}%
\bibitem [{\citenamefont {Benn}\ and\ \citenamefont
  {Tucker}(1983)}]{Benn1983fermions}%
  \BibitemOpen
  \bibfield  {author} {\bibinfo {author} {\bibfnamefont {I.}~\bibnamefont
  {Benn}}\ and\ \bibinfo {author} {\bibfnamefont {R.}~\bibnamefont {Tucker}},\
  }\bibfield  {title} {\bibinfo {title} {Fermions without spinors},\
  }\href@noop {} {\bibfield  {journal} {\bibinfo  {journal} {Communications in
  Mathematical Physics}\ }\textbf {\bibinfo {volume} {89}},\ \bibinfo {pages}
  {341} (\bibinfo {year} {1983})}\BibitemShut {NoStop}%
\bibitem [{\citenamefont {Catterall}(2023)}]{SimonPRD}%
  \BibitemOpen
  \bibfield  {author} {\bibinfo {author} {\bibfnamefont {S.}~\bibnamefont
  {Catterall}},\ }\bibfield  {title} {\bibinfo {title} {'t hooft anomalies for
  staggered fermions},\ }\href {https://doi.org/10.1103/PhysRevD.107.014501}
  {\bibfield  {journal} {\bibinfo  {journal} {Phys. Rev. D}\ }\textbf {\bibinfo
  {volume} {107}},\ \bibinfo {pages} {014501} (\bibinfo {year}
  {2023})}\BibitemShut {NoStop}%
\bibitem [{\citenamefont {Susskind}(1976)}]{susskind1976}%
  \BibitemOpen
  \bibfield  {author} {\bibinfo {author} {\bibfnamefont {L.}~\bibnamefont
  {Susskind}},\ }\bibfield  {title} {\bibinfo {title} {Lattice fermions},\
  }\href@noop {} {\bibfield  {journal} {\bibinfo  {journal} {Physical Review
  D}\ }\textbf {\bibinfo {volume} {16}} (\bibinfo {year} {1976})}\BibitemShut
  {NoStop}%
\bibitem [{\citenamefont {{Bi}}\ \emph {et~al.}(2015)\citenamefont {{Bi}},
  \citenamefont {{Rasmussen}}, \citenamefont {{Slagle}},\ and\ \citenamefont
  {{Xu}}}]{Bi:2015qv}%
  \BibitemOpen
  \bibfield  {author} {\bibinfo {author} {\bibfnamefont {Z.}~\bibnamefont
  {{Bi}}}, \bibinfo {author} {\bibfnamefont {A.}~\bibnamefont {{Rasmussen}}},
  \bibinfo {author} {\bibfnamefont {K.}~\bibnamefont {{Slagle}}},\ and\
  \bibinfo {author} {\bibfnamefont {C.}~\bibnamefont {{Xu}}},\ }\bibfield
  {title} {\bibinfo {title} {{Classification and description of bosonic
  symmetry protected topological phases with semiclassical nonlinear sigma
  models}},\ }\href {https://doi.org/10.1103/PhysRevB.91.134404} {\bibfield
  {journal} {\bibinfo  {journal} {\prb}\ }\textbf {\bibinfo {volume} {91}},\
  \bibinfo {eid} {134404} (\bibinfo {year} {2015})},\ \Eprint
  {https://arxiv.org/abs/1309.0515} {arXiv:1309.0515 [cond-mat.str-el]}
  \BibitemShut {NoStop}%
\bibitem [{\citenamefont {Joos}(1982)}]{r9}%
  \BibitemOpen
  \bibfield  {author} {\bibinfo {author} {\bibfnamefont {P.~B. .~H.}\
  \bibnamefont {Joos}},\ }\bibfield  {title} {\bibinfo {title} {The
  dirac-kahler equation and fermions on the lattice},\ }\href@noop {}
  {\bibfield  {journal} {\bibinfo  {journal} {Zeitschrift f{\"u}r Physik C
  Particles and Fields}\ } (\bibinfo {year} {1982})}\BibitemShut {NoStop}%
\bibitem [{\citenamefont {Kruglov}(2001)}]{r12}%
  \BibitemOpen
  \bibfield  {author} {\bibinfo {author} {\bibfnamefont {S.}~\bibnamefont
  {Kruglov}},\ }\bibfield  {title} {\bibinfo {title} {Dirac-kahler equation},\
  }\href@noop {} {\bibfield  {journal} {\bibinfo  {journal}
  {arXiv:hep-th/0110060v1}\ } (\bibinfo {year} {2001})}\BibitemShut {NoStop}%
\bibitem [{Note1()}]{Note1}%
  \BibitemOpen
  \bibinfo {note} {A chain is the sum of several simplexes.}\BibitemShut
  {Stop}%
\bibitem [{\citenamefont {JUNJI~KATO}\ and\ \citenamefont {UCHIDA}(2004)}]{R0}%
  \BibitemOpen
  \bibfield  {author} {\bibinfo {author} {\bibfnamefont {N.~K.}\ \bibnamefont
  {JUNJI~KATO}}\ and\ \bibinfo {author} {\bibfnamefont {Y.}~\bibnamefont
  {UCHIDA}},\ }\bibfield  {title} {\bibinfo {title} {Twisted superspace for
  n=d=2 super bf and yang–mills with dirac–kÄhler fermion mechanism},\
  }\bibfield  {journal} {\bibinfo  {journal} {International Journal of Modern
  Physics A}\ }\textbf {\bibinfo {volume} {19}},\ \href
  {https://doi.org/10.1142/S0217751X0401763X} {10.1142/S0217751X0401763X}
  (\bibinfo {year} {2004})\BibitemShut {NoStop}%
\bibitem [{\citenamefont {Abanov}\ and\ \citenamefont {Wiegmann}(2000)}]{r13}%
  \BibitemOpen
  \bibfield  {author} {\bibinfo {author} {\bibfnamefont {A.}~\bibnamefont
  {Abanov}}\ and\ \bibinfo {author} {\bibfnamefont {P.}~\bibnamefont
  {Wiegmann}},\ }\bibfield  {title} {\bibinfo {title} {Theta-terms in nonlinear
  sigma-models},\ }\href
  {https://doi.org/https://doi.org/10.1016/S0550-3213(99)00820-2} {\bibfield
  {journal} {\bibinfo  {journal} {Nuclear Physics B}\ }\textbf {\bibinfo
  {volume} {570}},\ \bibinfo {pages} {685} (\bibinfo {year}
  {2000})}\BibitemShut {NoStop}%
\bibitem [{\citenamefont {Hatcher}()}]{r16}%
  \BibitemOpen
  \bibfield  {author} {\bibinfo {author} {\bibfnamefont {A.}~\bibnamefont
  {Hatcher}},\ }\href {https://pi.math.cornell.edu/~hatcher/AT/ATpage.html}
  {\emph {\bibinfo {title} {Algebraic Topology}}}\BibitemShut {NoStop}%
\end{thebibliography}%

\newpage
\newpage

\appendix
\section{Review of Continuum K\"ahler-Dirac Fermions}
\label{appendixA} In this section, we present a concise overview of the principle of K\"ahler-Dirac fermions and the fundamental notations of differential geometry. The K\"ahler-Dirac equation generalizes the Dirac equation and was originally proposed by K\"ahler, who demonstrated that fermions with spin 1/2 can be formulated from differential forms. To better comprehend the K\"ahler-Dirac equation, we need to establish some geometrical prerequisites.

Firstly, we introduce the notations of differential and dual-differential. The operator $d$ maps $p$-forms to $p+1$-forms. The Hodge dual of a k-form $dx^{I}=dx^{i_1}\wedge\cdots\wedge dx^{i_k}$ is an (n-k) form that corresponds to the complementary set $\overline{I}={\overline{i}_1<\cdots\overline{i}_{n-k}}$. This can be expressed as:

\begin{gather}
\star dx^{I}=sgn({\sigma(I)})tdx^{\overline{I}}
\end{gather}
where $sgn({\sigma(I)})$ is the sign of permutation $\sigma(I)=i_1\cdots i_k\overline{i}_1\cdots\overline{i}_{n-k}$ and $t=\Pi_i(dx^{i}\iota dx^{i})$. Using the identity that applying the Hodge star twice leaves a $p$ form unchanged up to a sign, i.e., $\star\star\omega^p=(-1)^{p(n-p)}\eta\omega^p$, where $\eta=1$ for Euclidean spacetime and $\eta=-1$ for Minkowski spacetime. We can define the inverse of Hodge star as $\star^{-1}=(-1)^{p(n-p)}\star$. The notation of Hodge dual helps us define the dot product on the lattice. The interior product of two $p$-forms $\omega$ and $\eta$ is given by $(\omega,\eta)=\int_{\mathcal{M}}\overline{\omega} \wedge \star \eta$, where $\overline{\omega}$ is complex conjugation of $\omega$. On manifolds with this interior product structure, we can find the adjoint operator of the differential operator, $(\omega,d\eta)=(\delta\omega,\eta)$, where $\delta=(-1)^p\star^{-1}d\star$ satisfies this property. Since the differential operator satisfies $d^2=0$, the dual differential also has the corresponding property $\delta^2=0$.

Now, we can define the K\"ahler-Dirac equation in continuum Euclidean space and time:

\begin{gather}
(d-\delta- m)\Phi=0
\end{gather}
where $\Phi$ denotes the superposition of different differential forms $\Phi=\sum_{I_H}\phi_{I_H}dx^{I_H}$, and $K=d-\delta$ is the K\"ahler operator. The K\"ahler-Dirac equation generalizes the Dirac equation since $K^2=(d-\delta)^2=-(d\delta+\delta d)$ is the Laplace operator on manifolds. And $(d-\delta)^\dagger=-(d-\delta)$ is an anti-symmetric operator which can be decomposed into $\partial_{\mu}\gamma^{\mu}$, where $\gamma^{\mu}$ are $d+1$ mutually anti-commuting gamma matrices .

Moreover, the K\"ahler-Dirac equation can be written using the notation of Clifford product:
\begin{gather}
(\partial_{\mu}\vee^-dx^{\mu}-m)\Phi=0\qquad\mu=1,2\cdots, d+1
\end{gather}
 where $\vee^{-}=\wedge-\iota$ denotes Clifford product on vector space and $\iota$ is the interior product, where $dx^{\mu}\iota dx^{\nu}=\eta^{\mu\nu}$. If we represent Clifford product in the space of diferential forms, we can get the following corresponding relation $\vee^-dx^{\mu}\rightarrow\gamma^{\mu}$, where $\gamma^{\mu}$ satisfy $\{\gamma^{\mu},\gamma^{\nu}\}=2\eta^{\mu\nu}$ is representation of Clifford algebra $\mathcal{C}l(d+1)$ in $d+1$ dimensional flat spacetime.

It's starightforward to get the Kaher-Dirac equation in the presence of gauge field by replaced the derivative $\partial^{\mu}$ by covariant derivative $\partial^{\mu}-ieA^{\mu}$
\begin{gather}
((\partial_{\mu}-ieA_{\mu})\vee^-dx^{\mu}-m)\Phi=0\qquad\mu=1,2\cdots, d+1
\end{gather}
\section{Domain wall Fermions of K\"ahler-Dirac Fermions}
\label{appendixB}To discuss the anomaly inflow of K\"ahler-Dirac fermions, we need to indentify what the boundary sates of K\"ahler-Dirac fermions look like. We consider domain wall fermion in $d+1$ dimension defined on manifold $M^{d-1}\times R$ with coordinates $(x^a,z),a=1,2\cdots d-1$. Mass term $M(z)$ is function of $z$ and changes it sign at $z=0$ and we except massless fermions would appear on the boundary
\begin{gather}
    (\partial_t\vee^-dt+\partial_a\vee^-dx^a+\partial_z\vee^-dz-M(z))\psi(x^a,z,t)=0
\end{gather}
where $M(z)>0$ when $z>0$ and $M(z)<0$ when $z<0$ and we choose metric $\eta^{\mu\nu}=(-,+,\cdots,+)$.
$\psi$ can be decomposed into $\psi(x^a,z,t)=\chi(x^a,t)\wedge \phi(z)$, where $\chi(x^a)$ only contain forms without $dz$ and don't doesn't on $z$. And it gives:
\begin{gather}
    \partial_z\vee^-dz \phi(z)=M(z)\phi(z)\nonumber\\
   (\partial_t\vee^-dt+ \partial_a\vee^-dx^a)\chi(x^a)=0
 \end{gather}
since $(\vee^+dz)^2=1$ in Minkowski space and time and the eigenstate $i\vee^+dz$ with eigenvalue 1 is normalized state localized near the boundary. This equation described a exponentially localized domain wall fermion. And the freedoms the constraint of different forms on surface. And $\chi(x^a)$ satisfied the equation of motion of  massless $(d-1)+1$ dimensional K\"ahler-Dirac fermions.
\section{Continuum K\"ahler Ddirac fermions in $2D$}
\label{appendixC}
In this appendix, we present the explicit procedure for calculating the K\"ahler-Dirac model in $2D$. Any forms can be represented as $\Phi=\phi+\phi_1dx^1+\phi_2dx^2+\phi_{12}dx^1\wedge dx^2$. Moreover, $d\Phi=\partial_1\phi dx^1+\partial_2\phi dx^2+(\partial_1\phi_2-\partial_2\phi_1)dx^1\wedge dx^2$. In the basis $1, dx^2, dx^1, dx^1\wedge dx^2$, we can employ the vector $\left(\begin{array}{c} \phi \\ \phi_2\\ \phi_1 \\ \phi_{12}\\ \end{array}\right)$ to represent all combinations of forms. And $\star\Phi=\phi_{12}-\phi_1dx^2+\phi_2dx^1+\phi dx^1\wedge dx^2$ In the chosen basis, both $d$ and the Hodge star $\star$ can be expressed as:
\begin{align}
d=\left(
\begin{array}{cccc}
 0 & 0 & 0 & 0 \\
 \partial_2 & 0 & 0 & 0 \\
 \partial_1 & 0 & 0 & 0 \\
 0 & \partial_1 & -\partial_2 & 0 \\
\end{array}
\right),
\star=\left(
\begin{array}{cccc}
 0 & 0 & 0 & 1 \\
 0 & 0 & -1 & 0 \\
 0 & 1 & 0 & 0 \\
 1 & 0 & 0 & 0 \\
\end{array}
\right
)
\end{align}

By making use of the relation $\star^{-1}=(-1)^{p(n-p)}\eta\star$, we can demonstrate that $\delta=-\star d\star$ in the two-dimensional case. Upon performing some matrix calculations, we establish that $\delta$ is the Hermitian conjugation of $d$.
\begin{gather}
\delta=-\star d \star=\nonumber\\
-\left(
\begin{array}{cccc}
 0 & 0 & 0 & 1 \\
 0 & 0 & -1 & 0 \\
 0 & 1 & 0 & 0 \\
 1 & 0 & 0 & 0 \\
\end{array}
\right)
\left(
\begin{array}{cccc}
 0 & 0 & 0 & 0 \\
 \partial_2 & 0 & 0 & 0 \\
 \partial_1 & 0 & 0 & 0 \\
 0 & \partial_1 & -\partial_2 & 0 \\
\end{array}
\right)
\left(
\begin{array}{cccc}
 0 & 0 & 0 & 1 \\
 0 & 0 & -1 & 0 \\
 0 & 1 & 0 & 0 \\
 1 & 0 & 0 & 0 \\
\end{array}
\right)\nonumber \\
=\left(
\begin{array}{cccc}
 0 & \partial_2 & \partial_1 & 0 \\
 0 & 0 & 0 & \partial_1 \\
 0 & 0 & 0 & -\partial_2 \\
 0 & 0 & 0 & 0 \\
\end{array}
\right)
\end{gather}

And then we can check whether $K$ and $\Tilde{K}$ can be decomposed into gamma matrixes.
\begin{align}
&K=(d-\delta)=\left( \begin{array}{cccc}
  0 & -\partial_2 & -\partial_1 & 0 \\
 \partial_2 & 0 & 0 & -\partial_1 \\
 \partial_1 & 0 & 0 & \partial_2 \\
 0 & \partial_1 & -\partial_2 & 0 \\
\end{array}
\right)=-\ii\partial_1\sigma^{20}-\ii\partial_2\sigma^{32}
\\
 &\Tilde{K}=d+\delta=\left(
\begin{array}{cccc}
  0 & \partial_2 & \partial_1 & 0 \\
 \partial_2 & 0 & 0 & \partial_1 \\
 \partial_1 & 0 & 0 & -\partial_2 \\
 0 & \partial_1 & -\partial_2 & 0 \\
\end{array}
\right)=\partial_1\sigma^{10}+ \partial_2\sigma^{31}
\end{align}
\begin{table*}[hbt]
\begin{center}
\caption{In the accompanying table, we delineate the manifestations of the geometric quantities $K$, $\Tilde{\gamma}$, and $\Gamma$ within our carefully selected basis of Euclidean spacetime. This basis has been chosen to facilitate the representation of all geometric quantities as tensor products of Pauli matrices. The specific purpose of this choice is to ensure that $K$ can be depicted as a tensor product of Pauli matrices, corroborated by confirming that $K$ has a real representation.The procedure to select this basis is rooted in a systematic approach, dividing the differential forms into two distinct categories. The initial categorization is based on the presence or absence of $dx^1$; forms that contain $dx^1$ comprise the second group, while those without are part of the first group. This process is iterated within each group, further segregating them into subgroups based on whether they contain $dx^2$. This procedure is replicated until the basis is unequivocally determined. }
\begin{ruledtabular}
\begin{tabular}{ccccc}
d+1 & K & $\Gamma$ & $\Tilde{\gamma_i}$ & Basis \\
\hline
1 & $-\ii\partial_1\sigma^2$ & $\sigma^3$ & $i\sigma^1$ & 1, $dx^1$ \\
2 & $-\ii\partial_1\sigma^{20}-\ii\partial_2\sigma^{32}$ & $\sigma^{33}$ & $i\sigma^{10}$,$i\sigma^{31}$ & 1,$dx^2$, $dx^1$, $dx^1\bigwedge dx^2$ \\
3 & $-\ii\partial_1\sigma^{200}-\ii\partial_2\sigma^{320}-\ii\partial_3\sigma^{332}$ & $\sigma^{333}$ & $i\sigma^{100}$,$i\sigma^{310}$,$i\sigma^{331}$ 
& 1, $dx^3$, $dx^2$, $dx^2\bigwedge dx^3$, $dx^1$, \\
&&&& $dx^1\bigwedge dx^3$ ,$dx^1\bigwedge dx^2$, $dx^1\bigwedge dx^2\bigwedge dx^3$
\\
4 & $-\ii\partial_1\sigma^{2000}-\ii\partial_2\sigma^{3200}-\ii\partial_3\sigma^{3320}-\ii\partial_4\sigma^{3332}$ & $\sigma^{3333}$ & $i\sigma^{1000}$, $i\sigma^{3100}$, $i\sigma^{3310}$, $i\sigma^{3331}$ & $\ldots$\\
$\ldots$&$\ldots$&$\ldots$&$\ldots$&$\ldots$
\end{tabular}
\end{ruledtabular}
\end{center}
\end{table*}
\section{Products on Lattice}
\label{appendix:pro}
Though we use the idea of exterior product and interior product on lattice, but in fact cap product and cup prodcut are mathematical  rigorous  definition  of products on lattice. We introduce these ideas in this appendix. More mathematic details can be found in Ref\cite{r16}. Using the rigorous definition, we can define K\"ahler-Dirac equation on arbitrary lattice whose co-chain is isomorphic to chain.

First, our field is grassmanian number value co-chain rather than simplex on hypercubic lattice, which maps a cell to a grassmanian number. But the co-lattice and lattice is isomorphic for hypercubic lattice, so we don't bother distinguish it in the above discussion. On the hypercubic lattice, the co-chain is isomorphic to chain, and we denoted a basis of co-chain as $\omega^{\bm{\mu}}(x)$ as the dual of cell $e^{\bm{\mu}}(x)$ which satisfied $\omega^{\bm{\mu}}(x)(e_{\bm{\mu}'}(x'))=\delta_{x,x'}\delta_{\bm{\mu},\bm{\mu}'}$
The cup product $\cup$ is the generalization of the exterior product $\wedge$ acting on co chain, and is determined by the following equation:

\begin{gather}
\omega^{x,\bm{\mu}}\cup \omega^{x';\bm{\mu}'}=(-1)^{\bm{\mu}\cup \bm{\mu}'}\delta^{x,x'}\omega^{x;\bm{\mu}\cup \bm{\mu}'}
\end{gather}
 where $(-1)^{\bm{\mu}\cup \bm{\mu}'}$ represents the sign of permutation $(\mu_1,\mu_2,\cdots,\mu'_1,\mu'_2\cdots)$, with $\mu_1=(\mu_1<\mu_2\cdots<\mu_p)$,$\mu_1<\mu_2\cdots<\mu_p$ and $\bm{\mu}_2=(\mu'_1,\mu'_2\cdots)$, $\mu_1'<\mu_2'\cdots<\mu_p'$.

Although the cap product is initially defined between co-chain and chain, the application of Pontrjagin duality permits the definition of the interior product of co-chain $\omega^{x;\mu_1\cdots \mu_k}\cap \omega^{x';\mu_1'\cdots \mu'_k}=\eta^{\mu_1,\mu_1'}\cdots\eta^{\mu_p,\mu_p'}\delta^x_{x'}$. The cap product $\cap$ can be considered as the Hermitian conjugation of the cup product $\cup$, adhering to the relation $(\omega_1,\rho\cup\omega_2)=(\rho\cap\omega_1,\omega_2)$:

\begin{gather}
\omega^{x;\bm{\mu}}\cap\omega^{x';\bm{\mu}'}=(-1)^{\bm{\mu}/ \bm{\mu}'}\delta^{x,x'}\omega^{x;\bm{\mu}/\bm{\mu}'}
\end{gather}
By introducing the concepts of cap product $\cap$ and cup product $\cup$, the lattice K\"ahler-Dirac fermions can be further simplified and made analogous to the conventional definition of continuum K\"ahler-Dirac fermions through Clifford product. We shall not delve into the mathematical details of products on co-chain but will only demonstrate how to compute them on a square lattice. More comprehensive information regarding algebraic topology. Utilizing the notation, above we can reformulate the K\"ahler-Dirac equation on a square lattice as:
\begin{gather}
(\partial_{\mu}^-e^{\mu}\cup+\partial_{\mu}^+ e^{\mu}\cap-m)\Phi(x)=0
\end{gather}
where $e^{\mu}=\sum_{x}e^{x;\mu}$ is summation of 1 co-chain on the square lattice. 

Deriving the continuum K\"ahler-Dirac fermions from the aforementioned equation is rather straightforward. In the continuum model, there is no distinction between $\partial^{\pm,\mu}$, and the 1 co-chain $e^{\mu}$ should be replaced by the 1-form $dx^{\mu}$:

\begin{equation}
(\partial_{\mu}dx^{\mu}(\wedge-\iota)-m)\Phi(x)=0
\end{equation}

Here, the exterior product $\wedge$ and interior product $\iota$ are the direct analogs of cup product $\cup$ and cap product $\cap$, respectively. $\Phi$ is the superposition of distinct differential forms, denoted as $\Phi(x)=\sum_{I_H}\phi_{I_H}dx^{I_H}$. Furthermore, we consistently represent $\wedge-\iota$ as the Clifford product $\vee^-$.
 
\begin{table}[htb]
\label{tabel:2}
\caption{In the following table, we present a dictionary between operators on manifolds and their counterparts on lattices. Our analysis reveals a one-to-one correspondence between these operators, demonstrating that our discussion can be equally applicable to both lattice and manifold models in subsequent sections of this paper.}
\begin{center}
\begin{tabular}{cc}
\hline
    Continuum  or hypercubic lattice&General  lattice\\
    \hline
     Tensor $\mathcal{T}_V(p,0)$ & p chain $C_p$\\
     Hodge star $\star$ &  Hodge star $\star$ \\ 
     differential form $\Lambda^p$& co-chain (cells) $C^p$ \\
     exterior product $\wedge$ &cup product $\cup$ \\
     Interior product $\iota$ &cap product $\cap$ \\
     exterior differential d& co-boundary d\\
     dual differential $\delta$ & dual co-boundary $\delta$ \\
     partial $\partial_{\mu}$ & backward/forward partial $\partial^{\pm}_{\mu}$\\
 \hline
\end{tabular}
\end{center}
\end{table}

\end{document}